\documentclass[manuscript]{aastex}

\slugcomment{}

\shorttitle{Spectroscopy of Edge-on LSB Galaxies}
\shortauthors{Du Wei et al.}

\begin{document}

%% LaTeX will automatically break titles if they run longer than
%% one line. However, you may use \\ to force a line break if
%% you desire.

\title{Long-slit Spectroscopy of Edge-on Low Surface Brightness Galaxies}

%% Use \author, \affil, and the \and command to format
%% author and affiliation information.
%% Note that \email has replaced the old \authoremail command
%% from AASTeX v4.0. You can use \email to mark an email address
%% anywhere in the paper, not just in the front matter.
%% As in the title, use \\ to force line breaks.
\author{
Wei\ Du\altaffilmark{1,2},
Hong\ Wu\altaffilmark{1},
Yinan\ Zhu\altaffilmark{1},
WeiKang Zheng\altaffilmark{3}, and
Alexei V. Filippenko\altaffilmark{3}}
\affil{$^1$ Key Laboratory of Optical Astronomy, National Astronomical Observatories, Chinese Academy of Sciences, 20A Datun Road, Chaoyang District, Beijing, 100012, China}
\affil{$^2$ email: {\texttt wdu@nao.cas.cn}}
\affil{$^3$ Department of Astronomy, University of California, Berkeley, CA 94720-3411, USA}

%% Notice that each of these authors has alternate affiliations, which
%% are identified by the \altaffilmark after each name.  Specify alternate
%% affiliation information with \altaffiltext, with one command per each
%% affiliation.

%% Mark off your abstract in the ``abstract'' environment. In the manuscript
%% style, abstract will output a Received/Accepted line after the
%% title and affiliation information. No date will appear since the author
%% does not have this information. The dates will be filled in by the
%% editorial office after submission.

\begin{abstract}
We present long-slit optical spectra of 12 edge-on low surface brightness galaxies (LSBGs) positioned along their major axes. After performing reddening corrections for the emission-line fluxes measured from the extracted integrated spectra, we measured the gas-phase metallicities of our LSBG sample using both the [N~{\sc{ii}}]/H$\alpha$ and the $R_{23}$ diagnostics. Both sets of oxygen abundances show good agreement with each other, giving a median value of 12 + log(O/H) = 8.26 dex. In the luminosity-metallicity plot, our LSBG sample is consistent with the behavior of normal galaxies.  In the mass-metallicity diagram, our LSBG sample has lower metallicities for lower stellar mass, similar to normal galaxies. The stellar masses estimated from $z$-band luminosities are comparable to those of prominent spirals. In a plot of the gas mass fraction versus metallicity, our LSBG sample generally agrees with other samples in the high gas mass fraction space. Additionally, we have studied stellar populations of 3 LSBGs which have relatively reliable spectral continua and high signal-to-noise ratios, and qualitatively conclude that they have a potential dearth of stars with ages $<1$~Gyr instead of being dominated by stellar populations with ages $>1$~Gyr. Regarding the chemical evolution of our sample, the LSBG data appear to allow for up to 30\% metal loss, but we cannot completely rule out the closed-box model. Additionally, we find evidence that our galaxies retain up to about 3 times as much of their metals compared with dwarfs, consistent with metal retention being related to galaxy mass. In conclusion, our data support the view that LSBGs are probably just normal disk galaxies continuously extending to the low end of surface brightness.

\end{abstract}

%% Keywords should appear after the \end{abstract} command. The uncommented
%% example has been keyed in ApJ style. See the instructions to authors
%% for the journal to which you are submitting your paper to determine
%% what keyword punctuation is appropriate.


%% From the front matter, we move on to the body of the paper.
%% In the first two sections, notice the use of the natbib \citep
%% and \citet commands to identify citations.  The citations are
%% tied to the reference list via symbolic KEYs. The KEY corresponds
%% to the KEY in the \bibitem in the reference list below. We have
%% chosen the first three characters of the first author's name plus
%% the last two numeral of the year of publication as our KEY for
%% each reference.

%% Authors who wish to have the most important objects in their paper
%% linked in the electronic edition to a data center may do so by tagging
%% their objects with \objectname{} or \object{}.  Each macro takes the
%% object name as its required argument. The optional, square-bracket 
%% argument should be used in cases where the data center identification
%% differs from what is to be printed in the paper.  The text appearing 
%% in curly braces is what will appear in print in the published paper. 
%% If the object name is recognized by the data centers, it will be linked
%% in the electronic edition to the object data available at the data centers  
%%
%% Note that for sources with brackets in their names, e.g. [WEG2004] 14h-090,
%% the brackets must be escaped with backslashes when used in the first
%% square-bracket argument, for instance, \object[\[WEG2004\] 14h-090]{90}).
%%  Otherwise, LaTeX will issue an error. 

\section{Introduction}
Low surface brightness galaxies (LSBGs) were theorized to exist by \citet{Disney76} and then verified by the discovery of Malin~1 \citep{Bothun87}. They refer to galaxies which have lower surface brightness than the background \citep{Impey97} and are traditionally defined as galaxies with central surface brightness $\mu_{B}(0)$ fainter than some threshold which varies between 22.0 and 23.0 mag arcsec$^{-2}$ \citep[e.g.,][]{Impey01,Ceccarelli12}. Recent studies indicate that LSBGs contribute 20\% of the dynamical mass of galaxies in the Universe \citep{Minchin04} and 30\%--60\% of the number density of local galaxies \citep{McGaugh96,Bothun97,O'Neil00,Trachternach06,Haberzettl07}, suggesting that the contribution of LSBGs to the Universe can not be negligible. However, selection effects cause LSBGs to be underrepresented in the present-day galaxy catalogs. 

Compared with disk galaxies having high surface brightness (HSB), LSBGs generally tend to be dust-poor \citep[e.g.,][]{Matthews01} and H~{\sc{i}}-rich, with diffuse, low-density stellar disks and metallicity $\leq 1/3$ solar abundance \citep[e.g.,][]{McGaugh94}. This suggests that LSBGs have been inefficient star formers over their lifetimes and have low-level star-formation activity at present \citep[e.g.,][]{Das09,Galaz11}. Indeed, H~{\sc{i}} surface densities in LSBGs are often observed to be well below the critical threshold for star formation \citep[e.g.,][]{van der Hulst93}. Nonetheless, LSBGs typically contain some signatures of ongoing star formation, including modest amounts of H$\alpha$ emission from a very small number of H~{\sc{ii}} regions \citet[e.g.,][]{Schombert92, Schombert13} and blue colors indicative of young stellar populations \citep[e.g.,][]{McGaugh94_1, Matthews97}. 

Metallicities of stars and gas within a galaxy provide both a fossil record of its star formation history (SFH) and evolution. For galaxies, metallicities are traditionally represented by the oxygen abundance, 12 + log(O/H), which directly reflects the metal content of the gas in galaxies and indirectly reveals the metal content of the stars in galaxies that ionize the gas. For LSBGs, many previous works have demonstrated that LSBGs are metal-poor. The LSBG sample of \citet{McGaugh94} has been measured to have low oxygen abundances, 12 + log(O/H) $\leq 8.4$ dex. \citet{Burkholder01} found a mean value 12 + log(O/H) = 8.22 dex for 17 LSBGs. \citet{Lam15} studied KKR17, an LSBG dominated by H~{\sc{i}}, and measured an average 12 + log(O/H) = 8.0 dex. The oxygen abundances of \citet{de Naray04} are 12 + log(O/H) $\approx 7.8$--8.2 dex. Apparently, most LSBGs have 12 + log(O/H) lower than the solar value of oxygen abundance of 12 + log(O/H) = 8.87~dex \citep{Grevesse96}, and even lower than the transition value (12 + log (O/H) = 8.4~dex) between metal-poor and metal-rich branches for the $R_{23}$ method (see \S 4.3). To derive gas-phase metallicities of LSBGs, spectra should be required. However, it is not easy to derive spectra of LSBGs due to their low surface brightness. Fortunately, we have carried out spectroscopic observations for a sample of LSBGs. Later in this paper, we will introduce this observation and then study properties which are mainly relevant to metallicities on basis of the observational data.  

Given the faint nature of LSBGs, with peak surface brightness $\mu_{B}(0) \geq 22.5$ mag arcsec$^{-2}$, it is extremely difficult to obtain spectra of their stellar continua with relatively high signal-to-noise ratio ($> 10$) because the sky background always dominates the signal. Studies of the stellar populations of LSBGs are currently underrepresented and are based mainly on spectra of their H~{\sc{ii}} regions \citep[e.g.,][]{Schombert13} and central portions \citep[e.g.,][]{Morelli12}. But the bright H~{\sc{ii}} regions contain a large fraction of O- and B-type stars, so stellar populations studied through spectra of the H~{\sc{ii}} regions would be biased to younger ages. Compared to observing face-on LSBGs \citep[e.g.,]{Du15} or fiber spectroscopy, the spectral qualities can be highly improved by observing edge-on LSBGs, especially those having thin disks, with a single slit of appropriate length and width, positioned along the major axis. This scheme makes it possible, although still very difficult, to obtain reasonable-quality data. Very fortunately, spectra of several few LSBGs of our sample have relatively reliable stellar continua which can be used for stellar population studies, so we additionally study stellar populations of these few LSBGs and to investigate their star formation histories as byproducts of our observations (\S 6.1). 
%However, Additionally, the problem of dust attenuation can be minimized for edge-on LSBGs because LSBGs are generally poor in dust (although we find $A_V \approx 0.9$--1.0~mag for AGC253926).

%Numerous previous studies have shown that LSBGs are not abnormal objects, but rather are a ubiquitous product of disk-galaxy formation and evolution. However, as the contrast with respect to the background sky brightness is low, LSBGs are difficult to discover, and hence are easily missed and poorly understood. 

The outline of this paper is as follows.  Section 2 describes our sample, and detailed accounts of the observations and data reduction are given in \S 3 and \S 4. We study the physical properties of our edge-on LSBGs in \S 5, including oxygen abundances in \S 5.1, the luminosity-metallicity relation in \S 5.2, the mass-metallicity relation in \S 5.3, and the behavior of gas mass fraction versus metallicity in \S 5.4. In Section 6, we discuss stellar populations in \S 6.1, stellar mass estimation in \S 6.2, chemical evolution in \S 6.3, and the true nature of our LSBG sample in \S 6.4. Finally, we conclude in \S 7.

\section{The Sample}
A catalog of edge-on disk galaxies in SDSS DR7 \citep[EGIS;][]{Bizyaev14} consists of 5747 genuine edge-on galaxies in the Sloan Digital Sky Survey \citep[SDSS;][]{Abazajian09}. By analyzing photometric profiles in each of the $g$-, $r$-, and $i$-band images, the EGIS catalog provides estimates for structural parameters of the stellar disks of the galaxies, such as the disk scale height ($h$), radial scale length ($z_{0}$), and central surface brightness ($\mu_{0}$). 
To further study the EGIS catalog in combination with H~{\sc{i}} data, we correlated it with the $\alpha$.40 catalog \citep{Haynes11}, which lists H~{\sc{i}} line sources from 40\% of the area of the final blind 21~cm H~{\sc{i}} ALFALFA (Arecibo Legacy Fast Arecibo L-band Feed Array) survey for local galaxies. Using the transformation formula of \citet{Smith02}, we derived the central surface brightness in the $B$ band, $\mu_{0}(B)$, from those in the $g$ and $r$ bands. Then, according to a widely used definition of a LSBG, there are 287 galaxies in the EGIS-$\alpha$.40 sample that can be identified as LSBGs with $\mu_{0}(B) > 22.5$ mag arcsec$^{-2}$.

%{\it Among the EGIS-$\alpha$.40 LSBG sample, 182 galaxies have $\mu_{0}(B) > 23.0$ mag arcsec$^{-2}$, two of which are super-thin disk galaxies \citep{van der Kruit81, van der Kruit88, van der Kruit01, de Grijs98} whose axial ratios between the disk scale length ($h$) and the scale height ($z_{0}$) in the $g$ band are larger than 10. Most of them are optically bulgeless, with bulge-to-total luminosity ratio $B/T < 0.1$ in the SDSS $r$ band.}

\begin{figure*}
 \centering
 \includegraphics[scale=0.45]{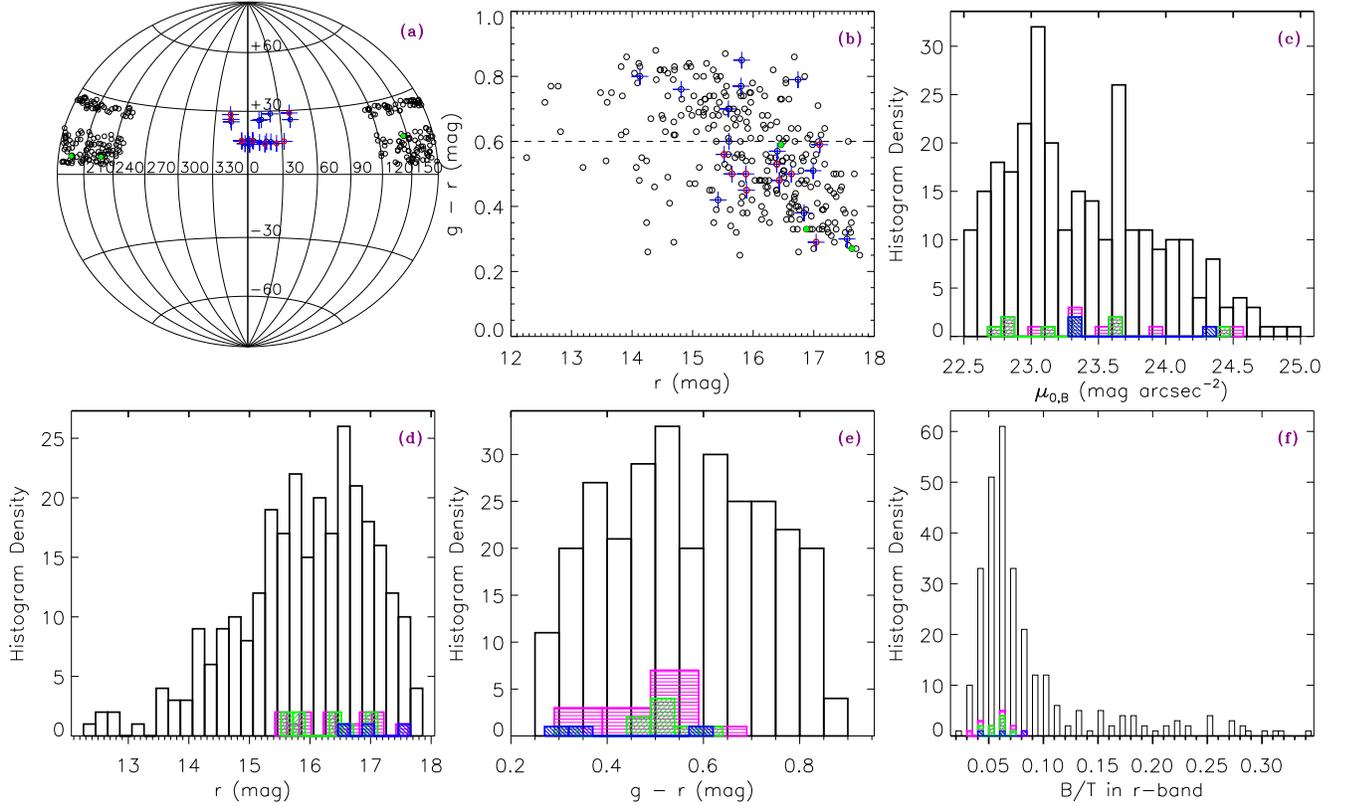}%targets_info.pro
 \caption[]{Main property distributions of 12 edge-on LSBGs in comparison with the total EGIS-$\alpha$.40  LSBG sample. In (a) and (b), blue plus signs are the 21 galaxies in the Fall sky region,  black open circles show the entire EGIS-$\alpha$.40 LSBG sample, red open circles indicate the final observed 9 galaxies, and green filled circles show the 3 edge-on LSBGs observed in the Spring sky during another program. In (c)--(f), the black, magenta with line fill, green with line fill, and blue with line fill respectively represent distributions of the entire EGIS-$\alpha$.40 LSBG sample, the prepared 14 LSBG targets, the final observed 9 LSBGs, and the 3 LSBGs from another observational program. The dashed black line in (b) shows $g-r = 0.6$~mag.\label{fig:target_info}}
\end{figure*}

Out of the EGIS-$\alpha$.40 LSBG sample, only 21 galaxies (blue filled circles in Figure ~\ref{fig:target_info}(a),(b)) are in the Fall sky region ($22.5^h < \alpha < 03^h$) which matched our available observing time. Of the 21 galaxies, 14 galaxies have $g-r < 0.6$~mag (dashed black line in Figure ~\ref{fig:target_info}), sufficiently blue to be considered as the final targets for our 3 nights in late-September 2015. Unfortunately, only 50\% of the total time was useful for observations because of stormy and cloudy conditions; we ultimately observed 9 galaxies. Comparing with the 14 prepared targets (magenta lines), the observed 9 galaxies (red) are generally consistent in distributions of $\mu_{0}(B)$, magnitude, color, and $B/T$ as shown in Figure ~\ref{fig:target_info}(c)--(f). Furthermore, comparing with the entire EGIS-$\alpha$.40 LSBG sample (black), these 9 galaxies (red) sample almost the full range in $\mu_{0}(B)$ from 22.5 to 24.5 mag arcsec$^{-2}$, but they sample only the fainter part in $r$ ($\gtrsim 15.5$~mag) and only the blue part in $g-r$ ($<0.6$~mag). 

On the other hand, in February 2015, we had already spectroscopically observed in a similar way several galaxies (not confined to be LSBGs) in the EGIS-$\alpha$.40 catalog, but with a different telescope using spare time from another program. From these objects, there are 3 galaxies with $\mu_{0}(B) > 22.5$ mag arcsec$^{-2}$. We overplot them (blue) in Figure ~\ref{fig:target_info} and find that they are comparable to our observed (Fall 2015) 9 LSBGs in distributions of main properties. So, we add these 3 LSBGs into our 9-galaxy LSBG sample in order to enlarge the overall observed sample in this paper.

\begin{figure*}
 \centering
 \includegraphics[scale=0.5]{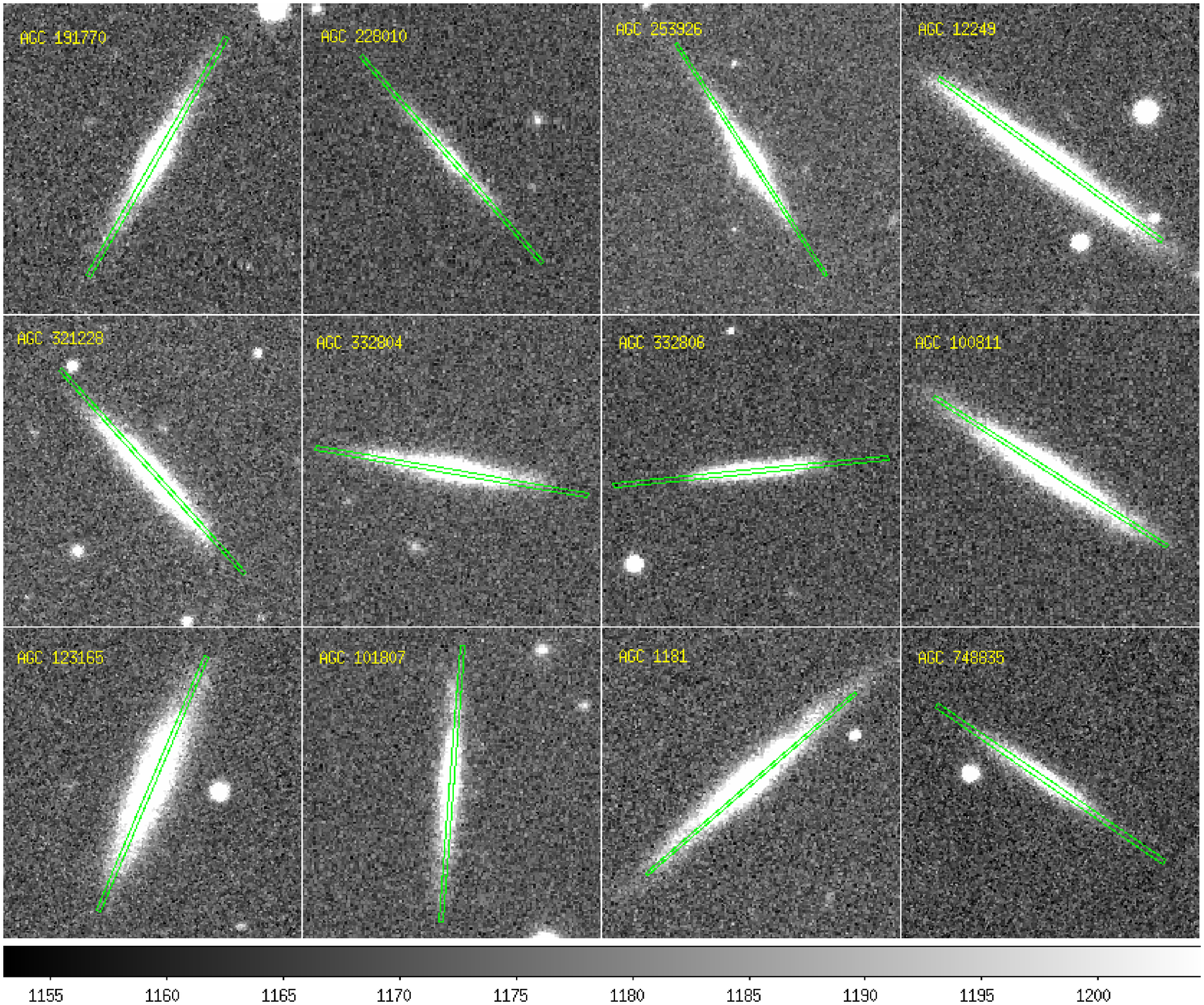}
 \caption[]{Schematic diagram of the positioning of the long slit when observing our 12 edge-on LSBGs.\label{fig:edgeon-slit}. In each panel, north is at top and east is to the left. The slit is indicated with a narrow green box which is scaled to the size of the slit of width 2$\arcsec$ and length 128$\arcsec$.  }
\end{figure*}

We display the SDSS images of the 12 edge-on LSBGs in Figure ~\ref{fig:edgeon-slit} and list their main properties in Table ~\ref{tbl-1}. Statistically, $\mu_{0}(B)$ values of the 12 LSBGs range from 22.6 to 24.4 mag arcsec$^{-2}$. There is one object (AGC 228010; see Figure ~\ref{fig:edgeon-slit}) with $h/z = 10.08$; it is a super-thin disk galaxy, previously defined to have an axial ratio between the disk scale length ($h$) and the scale height ($z_{0}$) in the $g$ band larger than 10.0 \citep{van der Kruit81, van der Kruit88, van der Kruit01, de Grijs98}. Additionally, our 12 galaxies have little (or no) optical bulges or central light concentrations, with $B/T < 0.08$ in the SDSS $r$ band, but they are not the only ones with $B/T < 0.08$ (see Figure ~\ref{fig:target_info}(f)).

%%%%%%%
%\begin{figure*}
% \centering
% \includegraphics[scale=0.40]{fig1.eps}%fig_images.pro
% \caption[]{SDSS $g$-, $r$-, and $i$-band combined images of 12 edge-on LSBGs from the EGIS-$\alpha$.40 sample that we selected for long-slit spectroscopy. Each panel has a size of $100'' \times 100''$. The pink triangles show stars for which there exist SDSS spectra, but they are otherwise of no interest here. \label{fig:sdss_images}}
%\end{figure*}
%%%%%%%%

\begin{deluxetable}{ccrrccrrrc}
\tabletypesize{\scriptsize}
\tablecaption{Properties of Edge-on LSBGs Taken from the EGIS-$\alpha$.40 Sample\label{tbl-1}}
\tablewidth{0pt}
\tablehead{
\colhead{Name} & \colhead{$\alpha$(J2000)} & \colhead{$\delta$(J2000)} & \colhead{$v_{\rm helio}$\tablenotemark{a}} & \colhead{Distance} &
\colhead{$\mu_{0}(B)$\tablenotemark{b}} & \colhead{$g$} & \colhead{$r$} & \colhead{$h$} & \colhead{$z$} \\

\colhead { } & \colhead{(deg)} & \colhead{(deg)} &\colhead{(km~s$^{-1}$)} &\colhead{(Mpc)} & \colhead{(mag arcsec$^{-2}$)}&\colhead{(mag)} & \colhead{(mag)}& \colhead{(arcsec)} &\colhead{(arcsec)}  \\

\colhead{  } &\colhead{PA} &\colhead{$B/T$\tablenotemark{c}}& \colhead{log($M$(H~{\sc{i}})/M$_\odot$)}&\colhead{$W_{50}$} & \colhead{$S_{21}$}  &\colhead{ } &\colhead{ }&\colhead{ }&\colhead{ } \\

\colhead{  }  & \colhead{(deg)} &\colhead{  } &  & \colhead{(km~s$^{-1}$)} & \colhead{(Jy km~s$^{-1}$)} &\colhead{ } &\colhead{ }&\colhead{ }&\colhead{ }
}
\startdata
191770&145.940&14.680&3823&57.4&23.27&17.21&16.88&6.04&1.60\\
      &149.97&0.06&9.07&165&1.50\\
\hline
228010&194.909&6.446&6354&95.3&24.31&17.90&17.63&10.39&0.96\\
      &41.12&0.04&9.45&165&1.32\\
\hline
253926&227.023&6.863&9333&135.9&23.33&17.05&16.46&6.55&1.55\\
      &32.85&0.08&10.26&264&4.19\\
\hline
12249&343.834&28.346&7549&102.9&23.59&16.08&15.52&11.87&1.93\\
     &53.80&0.05&9.73&295&2.15\\
\hline
321228&344.597&26.028&7492&102.1&22.84&16.92&16.39&5.92&1.71\\
     &42.00&0.06&9.26&104&0.75\\
\hline
332804&354.608&15.941&12425&172.3&23.16&16.91&16.43&6.85&1.65\\
      &80.07&0.06&9.99&266&1.40\\
332806&354.807&15.697&12055&167.0&22.88&17.33&17.04&4.24&1.05\\
      &95.95&0.05&9.83&242&1.03\\
\hline
100811&3.981&16.004&8362&114.4&23.67&16.38&15.88&10.44&2.12\\
      &57.34&0.06&9.77&239&1.89\\
\hline
123165&39.174&28.784&4536&61.9&22.68&16.1500&15.65&7.68&2.83\\
      &157.24&0.07&8.95&205&0.99\\
\hline
101807&14.957&14.328&11916&165.6&24.40&17.13&16.63&9.37&1.43\\
     &175.54&0.04&10.1&308&1.97\\
\hline
1181&25.125&14.523&8133&112.0&23.31&16.34&15.89&10.19&2.20\\
    &130.95&0.06&9.93&292&2.89\\
\hline
748835&31.840&15.429&12793&178.9&23.37&17.69&17.10&5.41&1.20\\
    &55.53&0.04&9.96&266&1.22\\
 \enddata
%% Text for table notes should follow after the \enddata but before
%% the \end{deluxetable}. Make sure there is at least one \tablenotemark
%% in the table for each \tablenotetext.
%\tablecomments{Table \ref{tbl-1} is published in its entirety in the 
%electronic edition of the {\it Astrophysical Journal}.  A portion is 
%shown here for guidance regarding its form and content.}
\tablenotetext{a}{Radial velocity from \citet{Haynes11}.}
\tablenotetext{b}{$B$-band central surface brightness from \citet{Bizyaev14}.}
\tablenotetext{c}{Bulge-to-total luminosity ratio of the galaxy from the SDSS catalog.}
\end{deluxetable}

\section{Observations}
\subsection{Overview}
Long-slit spectroscopy of three LSBGs in the Spring sky were obtained on 2015 Feb. 18 UT with the Kast double spectrograph \citep{Miller93} at the Cassegrain focus of the 3-m Shane telescope at Lick Observatory. A long slit 2$\arcsec$ wide was positioned along the major axis of each target, and the seeing was $\sim1.5\arcsec$. For the blue side, a 452 lines mm$^{-1}$ grating blazed at 3306~\AA\ was used in first order, yielding a wavelength range of 3150--6300~\AA\ with a spectral resolution of 1.41~\AA~pixel$^{-1}$. Use of a dichroic allows simultaneous observations at longer wavelengths with the red side; however, Lick spectroscopy was obtained only with the blue side.

The other nine LSBGs were observed on 2015 Sep. 21 UT (seeing $1.5\arcsec$) and Sep. 23 UT (seeing $1.2\arcsec$) with the 5-m Hale telescope at Palomar Observatory. We used the Double Spectrograph \citep[DBSP;][]{okegunn82} and a $128\arcsec$-long slit of width 2$\arcsec$, positioned along the major axis of each target. A D68 dichroic was used to divide the wavelength coverage into two regions, 3671--6727~\AA\ for the blue side (600 lines mm$^{-1}$ grating blazed at 4000~\AA) and 6223--9577~\AA\ for the red side (600 lines mm$^{-1}$ grating blazed at 10,000 \AA). The respective spectral resolutions are 1.078~\AA\ pixel$^{-1}$ and 0.819~\AA\ pixel$^{-1}$, so both gratings resolve all possible line features of interest to us. 

The total exposure times at Lick and Palomar for a galaxy were 3600--7200~s, subdivided into 2--4 subexposures in order to remove cosmic rays and enhance the signal-to-noise ratio (S/N) of the combined spectra. See Table ~\ref{tbl-2} for a log of observations.

\subsection{Light-Loss Problem}

The atmosphere is a dispersive medium. At nonzero zenith angles, a stellar image will be spread out into a small spectrum, and this spread can be significant compared to the size of the seeing disk. Under these conditions, if the spectrograph slit is not oriented along the direction of atmospheric dispersion (the parallactic angle), starlight at some wavelengths will not enter the narrow slit, while starlight at other wavelengths will pass into the instrument relatively unhindered.

During our observations, the slit was actually oriented along the major axis of the target object (position angle) instead of the parallactic angle, and the object was centered in the slit at $\sim6000$~\AA. Therefore, at nonzero zenith angles, object light from either wavelength end (blue or red) may be selectively lost. For example, as seen in Table ~\ref{tbl-2}, the spectrum of AGC 191770 at airmass 1.08 was obtained with  a 128$\arcsec \times$ 2$\arcsec$ slit on a night when the seeing was 1.5$\arcsec$. Then, following the calculations of \citet{Filippenko82}, if AGC 191770 was centered in the slit at a wavelength of 6000~\AA, it would be off center by 0.61$\arcsec$ at 3600~\AA\ $(R_{\lambda3600})$ and by 0.05$\arcsec$ in the opposite direction at 6500~\AA\ $(R_{\lambda6500})$. The parallactic angle was $3.05^\circ$ east of north, but the slit was actually aligned along a position angle of $149.97^\circ$ east of north. So, if AGC 191770 was a star with a Gaussian point-spread function, the percentage of light entering the slit at 3600~\AA\ $(T_{\lambda3600})$, 6000~\AA\ $(T_{\lambda6000})$, and 6500~\AA\ $(T_{\lambda6500})$ would be (respectively) 83.5, 88.3, and 88.3. Thus, the light loss at 3600~\AA\ would be 4.8\% greater than that at 6000~\AA.

Differential losses are a bigger concern for point sources than for extended objects such as galaxies. For extended sources, especially if the surface brightness is roughly constant over an area larger than the slit width, blue-end light loss from the position being observed can be partly gained back from neighboring positions within the source. The effect on the relative flux calibration will not be very large for our targets because of the following reasons. (1) Most spectra of our targets were observed at small airmasses (Table ~\ref{tbl-2}). (2) Our targets are edge-on galaxies with extended areas larger than the slit width (see Figure ~\ref{fig:edgeon-slit}), so the blue and red light losses are partly gained back from the neighboring positions within the objects. For completeness, we do the calculations above for all our targets (at the midpoint of the exposures) and list the results in Table ~\ref{tbl-2}.

\begin{deluxetable}{cccccccc}
\tabletypesize{\scriptsize}
\tablecaption{Log of Observations of Edge-on LSBGs \label{tbl-2}}
\tablewidth{0pt}
\tablehead{
\colhead{Name} & \colhead{UT Date} & \colhead{Exp.} & \colhead{blue $\lambda$ range} & \colhead{red $\lambda$ range} & \colhead{Tel. \& Instr.} &\colhead{airmass}  & \colhead{Position Angle} \\

\colhead {} & \colhead{YYYY-MM-DD} & \colhead{(s)} & \colhead{(\AA)} & \colhead{(\AA)}& \colhead{} &\colhead{} &\colhead{deg (East of North)} \\

\colhead{ } & \colhead{Parallactic Angle}  &\colhead{$R_{\lambda3600}$}&\colhead{$R_{\lambda6000}$}&\colhead{$R_{\lambda6500}$} &\colhead{$T_{\lambda3600}$} &\colhead{$T_{\lambda6000}$}&\colhead{$T_{\lambda6500}$}\\
\colhead{ } &\colhead{deg (East of North)} &\colhead{arcsec} &\colhead{arcsec}&\colhead{arcsec}&\colhead{$\%$ } &\colhead{$\%$ } &\colhead{$\%$ }\\

}
\startdata
191770&2015-02-18 & $2\times2700$ & 3450--6450 &	& Shane Kast &1.080 &149.97 \\
            &3.05 &0.61 &0.00 &-0.05 &83.5 &88.3 &88.3\\
\hline
228010&2015-02-18 & $2\times3600$ & 3450--6450 &	& Shane Kast &1.170 &41.12 \\
            &143.88 &0.80  &0.00 &-0.06 & 63.0  &88.3 &88.2\\
\hline
253926&2015-02-18 & $2\times2400$ & 3450--6450 &	& Shane Kast &1.170 &32.85 \\
            &24.11 &0.80 &0.00 & -0.06    &87.7 &88.3 & 88.3\\
\hline
12249 &2015-09-21 & $3\times1200$ & 3671--6727 &6223--9577 & Hale DBSP &1.239 &53.80 \\
           &108.52 &0.91 &0.00 &-0.08   &65.8 &95.0 &94.9\\
\hline
321228&2015-09-21 & $4\times1200$ & 3671--6727 &6223--9577& Hale DBSP &1.027 &42.00 \\
            &125.96 &0.97 &0.00 &-0.08   &52.7 &95.0 &94.8\\
\hline
332804&2015-09-21 & $3\times1200$ & 3671--6727 &6223--9577& Hale DBSP &1.048 &80.07 \\
            &177.59 &0.42 &0.00 & -0.03 &87.0 &95.0 & 94.9\\
\hline
332806&2015-09-21 & $4\times1200$ & 3671--6727 &6223--9577& Hale DBSP &1.055 &95.95 \\
            &18.51&0.47 &0.00 &-0.04  &85.6 &95.0 &94.9\\
\hline
100811&2015-09-21 & $4\times1200$ & 3671--6727 &6223--9577& Hale DBSP &1.117 &57.34 \\
            &45.41 &0.67 &0.00 & -0.05 &94.2 &95.0 &95.0\\
\hline
123165&2015-09-21 & $3\times1200$ & 3671--6727 &6223--9577& Hale DBSP &1.016 &157.24 \\
            &61.13 &0.27 &0.00 &-0.02  &91.9 &95.0 &95.0\\
\hline
101807&2015-09-23 & $3\times1200$ & 3671--6727 &6223--9577& Hale DBSP &1.057 &175.54 \\
            &0.38 &0.47 &0.00 &-0.04  &94.9 &95.0 &95.0\\
\hline
1181  &2015-09-23 & $3\times1200$ & 3671--6727 &6223--9577& Hale DBSP &1.062 &130.95  \\
          &16.35 &0.47 &0.00 &-0.04  &86.7 &95.0 &94.9\\
\hline
748835&2015-09-23 & $3\times1200$ & 3671--6727 &6223--9577& Hale DBSP &1.088 &55.53 \\
            &37.02 &0.57 &0.00 &-0.04  &93.6 &95.0 &95.0\\ 
\hline
\enddata
%% Text for table notes should follow after the \enddata but before
%% the \end{deluxetable}. Make sure there is at least one \tablenotemark
%% in the table for each \tablenotetext.
%\tablecomments{Table \ref{tbl-2} is published in its entirety in the 
%electronic edition of the {\it Astrophysical Journal}.  A portion is 
%shown here for guidance regarding its form and content.}
%\tablenotetext{a}{P.A. stands for Position Angle }
\end{deluxetable}

\section{Data Reduction}
\subsection{Overview}

Data reduction was carried out using standard IRAF procedures. After correction for overscan, bias,  and flat field, we reject cosmic rays using the IRAF task L.A.Cosmic \citep{van Dokkum01}, which identifies and removes cosmic rays robustly from astronomical images via a Laplacian algorithm. Then, we performed wavelength calibration on the two-dimensional (2D) spectral images of objects and standard stars. Sky subtraction will be described more detailedly in next subsection. After sky subtraction, we defined the appropriate aperture, traced the flux peaks along the dispersion direction, and interactively extracted one-dimensional (1D) spectrum by using the APALL task of IRAF. Finally, we averagely combined spectra from multiple exposures for an object into one final spectrum.

\begin{figure*}
 \centering
 \includegraphics[scale=0.11]{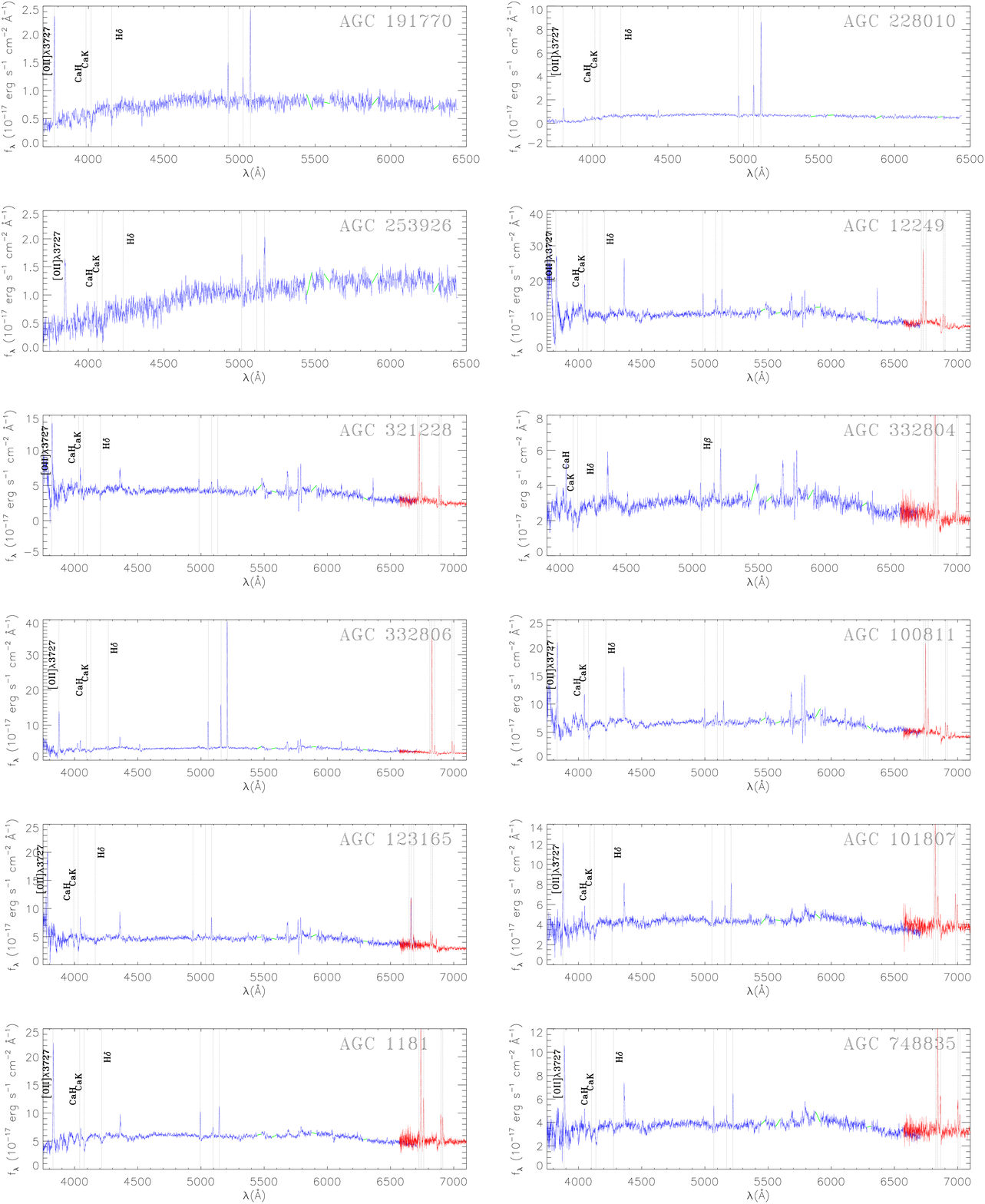}%fig_spectra_sed.pro(mac)
 \caption[]{Spectra of 12 edge-on LSBGs \label{fig:spectra2}. The blue- and red-side spectra from Lick (blue only; first 3 objects) and Palomar (next 9 objects) are shown in blue and red, respectively. The red-side spectrum is scaled to the flux level of the blue-side spectrum by using the wavelength region of overlap; the scale factors are $\sim1.2$. Here, we show only spectra at $\lambda > 3700$~\AA\ and (for the Palomar data) $\lambda < 7100$~\AA. This region contains the emission lines used in this study.}
\end{figure*}

In Figure ~\ref{fig:spectra2}, we show the combined blue-side spectra for the three galaxies observed with the Lick Shane telescope (they have no red-side observations). For the nine galaxies observed with the Palomar Hale telescope, both the blue- and red-side spectra are shown (with blue and red colors, respectively), after scaling the red side to match the flux level of the blue side in the small overlap region. Strong emission lines are marked. Generally, the night-sky lines are not clearly subtracted in the faint LSBG spectra, so we simply mask four wavelength regions having the strongest night-sky lines (green).

\subsection{Sky Subtraction}
Sky subtraction quality is very important for LSBGs as they are naturally fainter than the sky background in surface brightness. So, we describe the three methods of sky subtraction we have tried in data reduction for our observations below.

The first one is called the ``median-filter method.'' Before performing this method, we make the spatial and dispersion directions perpendicular to each other through rectification, such as by performing 2D wavelength calibration on spectral images of objects. Then, on the rectified images, assuming the direction of night-sky lines is along the column, we perform a median filter represented by a $1 \times n$ array along each column to generate a sky image. Here, $n$ must be larger than at least two times the width (major axis in pixels) of the object projected on the images. This method is good at removing sky lines but may overestimate the sky continuum blended in the object. 

The second one is the sky-fitting method adopted in the APALL task of IRAF, which uses a second-order polynomial to fit the background from the substantial sky region along the slit. This method can do well in estimating sky continuum blended in the object but does not do very well in removing sky lines. 

The third one is the \citet{Kelson03} method which makes full use of the raw image data (before the optical distortions and spectral line curvatures have been rectified) to model the sky background.  This method is not very good for our faint objects, because it either underestimates the sky continuum or leaves sky-line residuals in our images. 

None of the three methods above is sufficiently good at removing both sky continuum and sky lines clearly from spectra of our faint LSBGs. However, we have no other good ideas to implement it ideally. So, for Lick data which have no obvious distortions of raw images, the first ``median-filter" method has been used to simultaneously subtract the sky lines and sky continuum. However, for Hale data which have relatively serious distortions of raw images, this ``median-filter" method does not do well in sky continuum subtraction, so we finally used the second method to improve the sky continuum subtraction a little better, but we warn the reader that the final spectral continuum shapes for Hale spectra are still probably not accurate.

\subsection{Problems in the Spectra}   

Examining the extracted spectra, it appears that the reductions of the Hale spectra are not good at wavelengths $<3800$~\AA. Unexpectedly, many absorption lines appear in the observed spectra of our standard stars in this blue-end region due to unresolved reasons, so we are not able to determine an accurate continuum shape for fitting the flux-calibration curve in this blue-end region. Fortunately, except for AGC~123165, our LSBG galaxies all have redshift $z > 0.02$, so the bluest useful line ([O~{\sc{ii}}] $\lambda$3727) falls at $\lambda > 3800$~\AA. 
 
 Although we have extracted the red-side Palomar Hale spectra, their continuum S/N is quite low, so we abandoned using the red-side spectral continuum. However, the emission lines in the red spectra, such as [N~{\sc{ii}}] $\lambda$6584, H$\alpha$, and [S~{\sc{ii}}] $\lambda\lambda$6716, 6731, are reasonably strong, and they could therefore be used for dust-extinction corrections and oxygen-abundance measurements. 
 
 Moreover, as mentioned above, the spectral continuum shapes of the nine galaxies observed with the Hale telescope are not reliable, so we will not use them to draw any firm conclusions.

%% In this section, we use  the \subsection command to set off
%% a subsection.  \footnote is used to insert a footnote to the text.

%% Observe the use of the LaTeX \label
%% command after the \subsection to give a symbolic KEY to the
%% subsection for cross-referencing in a \ref command.
%% You can use LaTeX's \ref and \label commands to keep track of
%% cross-references to sections, equations, tables, and figures.
%% That way, if you change the order of any elements, LaTeX will
%% automatically renumber them.

%% This section also includes several of the displayed math environments
%% mentioned in the Author Guide.

\section{Results}
\subsection{Oxygen Abundance}

The gas-phase metallicity (oxygen abundance, $Z = 12 +$ log(O/H)) of a galaxy is a powerful probe of the past star formation and growth. It can also constrain the current stage of galaxy evolution as it is intimately connected to the fueling and regulation of star formation. Usually, flux ratios of strong optical emission lines are used as the abundance-sensitive diagnostics, such as $R_{23}$, [N~{\sc{ii}}]/[O~{\sc{ii}}], [N~{\sc{ii}}]/H$\alpha$, [N~{\sc{ii}}]/[S~{\sc{ii}}], and so on \citep{Pagel79,Kewley02}. Based on extracted spectra of our LSBGs, we measure the observed fluxes of strong emission lines in the blue-side ([O~{\sc{ii}}] $\lambda$3727, H$\beta$, [O~{\sc{iii}}] $\lambda\lambda$4959, 5007) and red-side ([N~{\sc{ii}}] $\lambda\lambda$6548, 6584, H$\alpha$, [S~{\sc{ii}}] $\lambda\lambda$6716, 6731). 

An intrinsic H$\alpha$/H$\beta$ flux ratio of 2.86 is assumed for Case B recombination at $10^{4}$~K \citep{Osterbrock89} because the relative intensities of the hydrogen Balmer lines are nearly independent of density and temperature. Therefore, for observed line fluxes measured from the 9 Hale spectra which have both blue- and red-side, we use the Balmer decrement (H$\alpha$/H$\beta$) and the CCM extinction law with $R_V = 3.1$ \citep{Cardelli89} to correct dust extinction. Based on these intrinsic fluxes, we compute several useful flux ratios of [N~{\sc{ii}}]/[O~{\sc{ii}}], [N~{\sc{ii}}]/[O~{\sc{ii}}], [N~{\sc{ii}}]/H$\alpha$, and $R_{23}$ [where $R_{23} =$ ([O~{\sc{ii}}] $\lambda$3727 + [O~{\sc{iii}}] $\lambda\lambda$4959, 5007)/H$\beta$], among which [N~{\sc{ii}}]/H$\alpha$ and $R_{23}$ are individually used as metallicity diagnostics. For the [N~{\sc{ii}}]/H$\alpha$ diagnostic, both the PP04 \citep{Pettini04} and D02 \citep{Denicolo02} calibrations are used to measure the oxygen abundances. For the $R_{23}$ diagnostic, both the M91 \citep{McGaugh91} and Z94 \citep{Zaritsky94} calibrations are employed to measure the oxygen abundances. Thus, we have four sets of oxygen abundances for each of the 9 Hale-observed galaxies except for AGC 123165 which have no $R_{23}$--based metallicities because its spectra has some problem in the blue-end wavelength range covering  [O~{\sc{ii}}] $\lambda$3727 (see \S 4.3). It should be noted that when using the M91 calibration, we need to determine which branch is appropriate for each galaxy because the $R_{23}$ versus O/H relation is double-valued \citep{Kewley02,de Naray04}. Distinguishing a branch (high-$Z$ or low-$Z$) can be accomplished with the [N~{\sc{ii}}] $\lambda$6584 line: strong (weak) nitrogen emission indicates that the high-$Z$ (low-$Z$) branch is appropriate. The dividing line is at log([N~{\sc{ii}}]/[O~{\sc{ii}}]) $\approx -1.2$ \citep{Kewley02}. However, for observed line fluxes measured from the 3 Lick spectra which have no observed red-side spectra (neither H$\alpha$ nor [N~{\sc{ii}}]), no reddening correction is performed and only $R_{23}$ diagnostic and $R_{23}$--based metallicities are measured. When using the M91 calibration, no available log([N~{\sc{ii}}]/[O~{\sc{ii}}]) can be used for distinguishing branches, so we empirically adopt the low-$Z$ branch to measure metallicities as LSBGs are generally acknowledged to be metal-poor. 

In Table ~\ref{tbl-3}, we report measurement results above and also show the average ($Z_{\rm ave1}$) of  both $R_{23}$-based metallicities ($Z_{\rm Z94}$ and $Z_{\rm M91}$) and the average ($Z_{\rm ave2}$) of both [N~{\sc{ii}}]/H$\alpha$-based metallicities ($Z_{\rm PP04}$ and $Z_{\rm D02}$) for each LSBG.
In statistic, our measurements show that the $R_{23}$-based metallicities ($Z_{\rm ave1}$) are 8.00--8.62 dex, with a mean of 8.30 dex and a median of 8.26 dex. The [N~{\sc{ii}}]/H$\alpha$-based metallicities ($Z_{\rm ave2}$) are 7.67--8.54 dex, with a mean of 8.20 dex and a median of 8.26 dex. Generally, the two sets of metallicities are consistent with each other and are both subsolar, with the $12 +$ log(O/H) value being 8.87 dex for the Sun \citep{Grevesse96}. Our metallicities for the LSBG sample agree with the results for LSBGs from \citet{de Naray04} (see Figure ~\ref{fig:lumi_metal}). 

  \begin{deluxetable}{ccccccccccc}
\tabletypesize{\scriptsize}
\tablecaption{Emission-Line Ratios\tablenotemark{a} and Oxygen Abundance of Edge-on LSBGs\label{tbl-3}}
\tablewidth{0pt}
\tablehead{
\colhead{AGC No.}  &\colhead{(H$\alpha$/H$\beta$)$_{\rm obs}$} &\colhead{log([N~{\sc{ii}}]/[O~{\sc{ii}}])} &\colhead{$R_{23}$}&\colhead{log([N~{\sc{ii}}]/H$\alpha$)} &\colhead{$Z_{\rm M91}$} &\colhead{$Z_{\rm Z94}$}&\colhead{$Z_{\rm PP04}$}&\colhead{$Z_{\rm D02}$}&\colhead{$Z_{\rm ave1}$\tablenotemark{b}}&\colhead{$Z_{\rm ave2}$\tablenotemark{c}}
}
\startdata
  191770  &         &          &5.93 &          &8.04    &8.68    &         &         &8.36   &  \\
  228010 &         &          &6.64   &          &7.80    &8.59      &         &          &8.195    &\\
  253926  &         &          &4.16   &          &7.78    &8.89      &          &          &8.335   &\\
  12249   &3.42  &-0.67  &5.98 &-0.70  &8.56   &8.67     &8.46    &8.61   &8.615  &8.535\\
  321228 &4.67  &-1.10  &11.44&-1.13  &8.00   &8.01     &8.22    &8.30    &8.005 &8.26\\
  332804 &5.79  &-1.30  &13.00 &-1.36  &8.79  &7.85  &8.14     &8.13    &8.32     &8.135\\
  332806  &3.53  &-0.98  &7.63   &-2.07  &8.53  &8.47    &7.72    &7.61  &8.5    &7.665\\
  100811  &4.19   &-0.92  &9.37   &-0.94  &8.22  &8.27   &8.31     &8.44   &8.245   &8.375\\
  123165 &6.24   &          &          &-1.08  &          &         &8.24     &8.33     &           &8.285\\
  101807  &4.55  &-1.05  &9.35   &-1.51   &8.24   &8.27  &8.08    &8.02   &8.255    &8.05\\
  1181     &5.30   &-1.01  &9.89  &-1.14   &8.15   &8.21   &8.22   &8.29   &8.18    &8.255\\
  748835 &4.76   &-1.04  &9.37  &-1.18   &8.20   &8.27   &8.20   &8.26    &8.235  &8.23\\
\hline
\enddata
%% Text for table notes should follow after the \enddata but before
%% the \end{deluxetable}. Make sure there is at least one \tablenotemark
%% in the table for each \tablenotetext.
%\tablecomments{Table \ref{tbl-2} is published in its entirety in the 
%electronic edition of the {\it Astrophysical Journal}.  A portion is 
%shown here for guidance regarding its form and content.}
\tablenotetext{a}{Corrected for reddening.}
\tablenotetext{a}{$Z_{\rm ave1}$ is the average value of $Z_{\rm M91}$ and $Z_{\rm Z94}$, which are both based on the $R_{23}$ diagnostic. }
\tablenotetext{b}{$Z_{\rm ave2}$ is the average value of $Z_{\rm PP04}$ and $Z_{\rm DO02}$, which are both based on the [N~{\sc{ii}}]/H$\alpha$ diagnostic.}
\end{deluxetable}

\subsection{The Luminosity-Metallicity Relation}

The strong correlation between galaxy luminosity ($L$) and galaxy metallicity ($Z$) is one of the most significant phenomenological results in chemical evolution studies of galaxies. Previous work found that the correlations between $L$ and $Z$ for irregular galaxies \citep{Lequeux79, Skillman89}, disk galaxies (predominantly high surface brightness galaxies, HSBGs; \citealt{Garnett87}), and elliptical galaxies \citep{Faber73,Brodie91} are very similar. Thus, it is instructive to plot our LSBG sample onto the familiar $L$--$Z$ relation defined by previous work on other galaxy types to validate the consistency of our measurements of LSBGs.

\begin{figure*}
 \centering
 \includegraphics[scale=0.50]{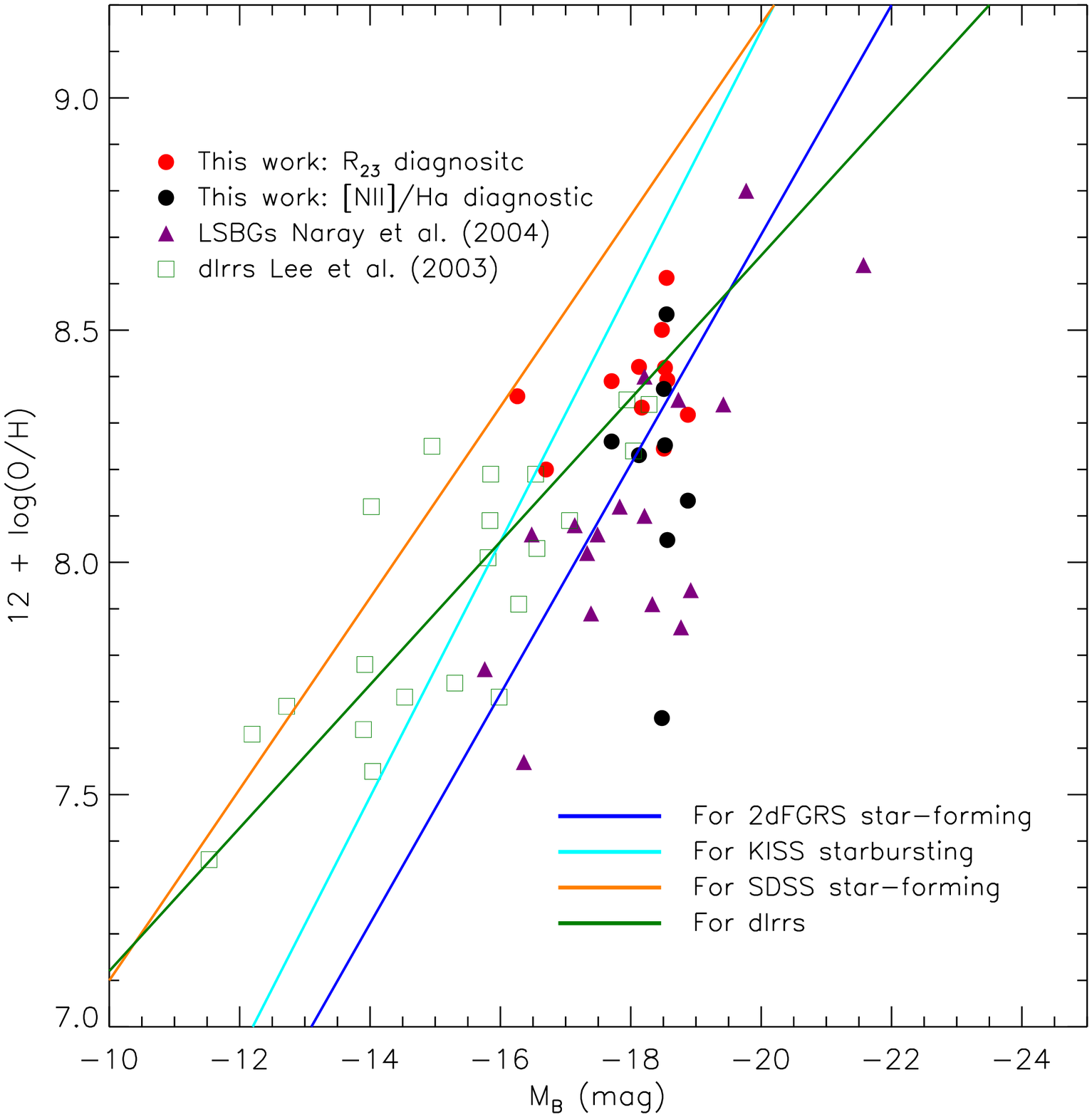}%fig_metal2.pro(linux)
 \caption[]{Luminosity-metallicity plot for our LSBG sample. The red and black filled circles respectively refer to the $R_{23}$-based and [N~{\sc{ii}}]/H$\alpha$-based metallicities. The purple triangles refer to the LSBG sample of \citet{de Naray04}. The green squares are the dwarf irregular galaxy sample of \citet{Lee03}. The blue solid line is the fitted $L$--$Z$ relation from \citet{Lamareille04}, which is determined for a large sample of 6387 star-forming galaxies in the local universe ($0 < z < 0.15$) from the 2dFGRS. The orange solid line is the relation determined by \citet{Tremonti04} for a large sample of 53,000 star-forming galaxies at $z \approx 0.1$ from the SDSS.  The cyan solid line is the relation of \citet{Melbourne02} for a large sample of $\sim900$ starbursting emission-line galaxies from KISS. The dark-green solid line is the relation for the dIrr sample (dark-green squares) of \citet{Lee03}. \label{fig:lumi_metal}}
\end{figure*}

In the $L$--$Z$ plot in Figure~\ref{fig:lumi_metal}, we show both $R_{23}$-based (red filled circles) and the [N~{\sc{ii}}]/H$\alpha$-based (black filled circles) metallicities for our LSBGs. Here, the $B$-band absolute magnitude for our LSBGs are computed from the galaxy distances (provided by $\alpha.$40 catalog) and the $B$-band apparent magnitudes (which are transformed from magnitudes of $g$- and $r$-bands measured by our own surface photometry \citep{Du15}). Data of LSBGs from \citet{de Naray04} (purple triangles) and dwarf irregulars (dIrrs; dark-green squares) from \citet{Lee03} are also plotted in this $L$--$Z$ figure for comparisons. Besides, we show a variety of recent fits to the $L$--$Z$ relation determined respectively for the dIrr sample of \citet{Lee03} (green line), a large sample of SDSS star-forming galaxies from \citet{Tremonti04} (orange line), a large sample of starbursting emission-line galaxies drawn from the KISS (KPNO International Spectroscopic Survey) sample from \citet{Melbourne02} (cyan line), and a large sample of 2dFGRS (Two-Degree Field Galaxy Redshift Survey) star-forming galaxies from \citet{Lamareille04} (blue line) from the literature. Among these relations, it seems that there is some disparities, which are probably caused by the utilization of different samples and different diagnostics and calibrations for measuring abundances.  

It is not easy to constrain a quantitatively reliable $L$--$Z$ relation for our LSBG data; they do not have sufficient leverage because of their small coverage in luminosity space. As shown in Figure ~\ref{fig:lumi_metal}, if all 12 galaxies (red filled circles) are considered, our sample covers only $\sim3$~mag of absolute magnitude, and if only the 9 galaxies (black filled circles) having both blue- and red-side spectra are considered, it covers only $\sim1$~mag. Instead of being quantitative, we thus carry out our studies qualitatively. From Figure ~\ref{fig:lumi_metal}, it appears that our LSBG sample with both sets of metallicities, especially with the [N~{\sc{ii}}]/H$\alpha$--based metallicities, seems to generally follow the similar distribution as the LSBGs from \citet{de Naray04} in their common covering region in luminosity and metallicity space. Both LSBG samples are consistent with the general distribution trend of other samples (normal galaxies, dIrrs) in this $L$--$Z$ plot, although with some degree of scatter.  

Additionally, we checked for quantitative consistency between our sample and the LSBG sample from \citet{de Naray04}. We did the test 50 times. Each time, we randomly picked a subsample of size equal to our small LSBG sample with [N~{\sc{ii}}]/H$\alpha$-based metallicities from the LSBG sample of \citet{de Naray04}, and then checked to see whether this subsample gives a distribution consistent with our small LSBG sample by using a K-S test. Our criteria for each K-S test to decide whether the subsample is consistent with being drawn from the same distribution as our small LSBG sample is the probability value being larger than 0.90 (P$>$0.90). The result is that in the total of 50 trials, 32\% of the subsamples from the \citet{de Naray04} LSBG sample give distributions that are consistent with our limited LSBG sample in the $L$--$Z$ plot. Any fraction more than 10\% and less than 90\% is basically inconclusive. So, such a fraction shows that the samples are too small to provide a conclusive result. If our sample could be improved to a larger size and wider luminosity coverage, such quantitative tests might give a more conclusive statistical fraction for consistency.

\subsection{The Mass-Metallicity Relation}
\subsubsection{Stellar Mass Estimate}
For our LSBGs, we derived two sets of stellar masses. One set is the ``$B$-based'' stellar masses which are derived by using their $B$-band luminosity and $B-V$ color transformed from  $g-r$ measured in \citep{Du15}. The other set is the ``$z$-based'' stellar masses in the SDSS MPA-JHU catalog which are computed from the SDSS $z$-band luminosities in synergy with the $z$-band stellar $M/L$ derived by the SDSS multi-band SED fitting method \citep{Kauffmann03}. Resultantly for our LSBGs, the ``$B$-based'' stellar masses are $M_{\star} \approx 10^{7.89}$--$10^{9.21}$ M$_{\odot}$, with a mean of $10^{8.81}$ M$_{\odot}$ and a median of $10^{9.09}$ M$_{\odot}$. The ``$z$-based'' stellar masses are $10^{8.37}$--$10^{9.96}$ M$_{\odot}$, with a mean of $10^{9.42}$ M$_{\odot}$ and a median of $10^{9.79}$ M$_{\odot}$. We report the results in Table ~\ref{tbl-4}.
Apparently, the set of ``$B$-based" stellar masses is on average $\sim0.6$ order of magnitude lower than the set of ``$z$-based" stellar masses. In this paper, we prefer the ``$z$-based" stellar masses for our LSBGs, and the reason will be given in \S 6.2. 

\begin{deluxetable}{ccccc}
\tabletypesize{\scriptsize}
\tablecaption{Estimated Stellar Mass and Gas Mass Fraction of Edge-on LSBGs \label{tbl-4}}
\tablewidth{0pt}
\tablehead{
\colhead{AGC No.}  & \colhead{log($M_{\star}$/M$_{\odot}$)$_{B-V}$} & \colhead{log($M_{\star}$/M$_{\odot}$)$_{\rm SDSS}$} &\colhead{$f_{g,B-V}$} &\colhead{$f_{g,{\rm SDSS}}$}
}
\startdata
  191770 &7.89 &8.39 &0.96    &0.87\\
  228010 &7.95 &8.37&0.98     &0.94\\
  253926 &9.11 &9.96 &0.95     &0.73\\
  12249   &9.21 &        &0.82     &      \\
  321228 &8.82 &        &0.79      &      \\
  332804  &9.20 &9.88&0.90      &0.64\\
  332806 &8.70 &9.46 &0.95    &0.76\\
  100811 &9.09 &9.79 &0.87     &0.57\\
  123165 &8.64 &        &0.74        &       \\
  101807 &9.11 &9.88  &0.93     &0.70\\
  1181    &9.00 &9.68   &0.92    &0.71\\
  748835 &9.09 &         &0.91     &       \\
\hline
\enddata
%% Text for table notes should follow after the \enddata but before
%% the \end{deluxetable}. Make sure there is at least one \tablenotemark
%% in the table for each \tablenotetext.
%\tablecomments{Table \ref{tbl-2} is published in its entirety in the 
%electronic edition of the {\it Astrophysical Journal}.  A portion is 
%shown here for guidance regarding its form and content.}
%\tablenotetext{a}{P.A. stands for Position Angle }
\end{deluxetable}

\subsubsection{$M$--$Z$ plot}
\begin{figure*}
 \centering
 \includegraphics[scale=0.60]{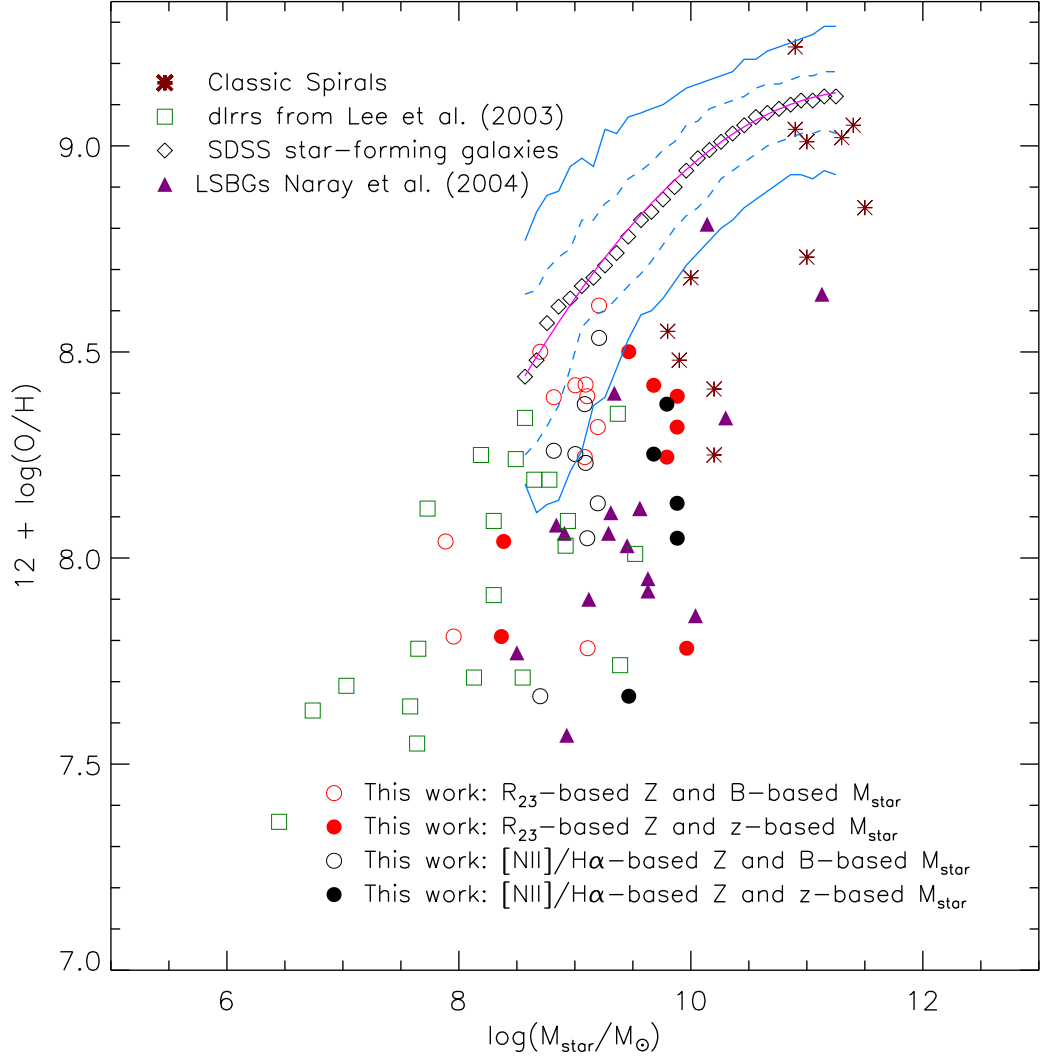}%fig_metal2.pro(Linux)
 \caption[]{Mass-metallicity plot for our LSBG sample. Our edge-on LSBGs are shown by red or black circles, respectively representing the $R_{23}$-based and [N~{\sc{ii}}]/H$\alpha$-based abundances.  The open and filled circles show stellar masses estimated respectively from the $B-V$ color and the SDSS SED-fitting. Purple triangles, brown asterisks, and dark-green squares stand for the LSBG sample from \citet{de Naray04}, the normal massive spiral galaxies from \citet{Garnett87}, and the dIrr sample from \citet{Lee03}. The black open diamonds indicate the medians of the 53,000 SDSS star-forming galaxies from \citet{Tremonti04} in bins of 0.1 dex in stellar mass, and the magenta line shows the polynomial fit to the median data from all of the bins. The cyan blue dashed and solid lines are respectively contours that enclose 68\% and 95\% of the 53,000 star-forming galaxies. \label{fig:mass_metal}}
\end{figure*}

We show our LSBG data in the mass-metallicity ($M$--$Z$) plot in Figure ~\ref{fig:mass_metal}. For metallicity, both $R_{23}$-based (black) and [N~{\sc{ii}}]/H$\alpha$-based (red) oxygen abundances are displayed. For stellar mass, both the ``$B$-based" (open circles) and ``$z$-based" (filled circles) stellar masses are shown. So, in this $M$--$Z$ plot, our LSBG sample appears as four individual sets of distributions, among which the two sets represented by red and black open circles are systematically biased to lower stellar mass because the ``$B$-based" stellar masses may be underestimated (as mentioned in \S 5.3.1 and discussed in \S 6.2). So, comparing with distributions represented by the red and black open circles, we prefer distributions based on the ``$z$-based" stellar mass estimations (red and black filled circles). For comparison, the LSBG sample (purple triangles) of \citet{de Naray04}, the dIrr sample (dark-green squares) of \citet{Lee03}, and the massive spiral galaxy sample (brown asterisks) of \citet{Garnett87} in the very local universe (including M101, M31, NGC~1365, M81, NGC~2997, and so on) are also included in this $M$--$Z$ plot. Additionally, the $M$--$Z$ relation determined for $\sim53,400$ SDSS star-forming galaxies of \citet{Tremonti04} is shown. The black open diamonds represent the median in bins of 0.1 dex in stellar mass that include at least 100 data points. The magenta line shows the polynomial fit of \citet{Tremonti04} to the median data. The cyan blue dashed and solid lines are the contours that enclose 68\% and 95\% of the data. 

Obviously, the distributions of our LSBG sample with ``$z$-based" stellar mass are generally consistent with the LSBG sample from \citet{de Naray04} at their common stellar mass space. Additionally, for normal star-forming galaxies and other types, there is a tendency that galaxies with lower stellar masses generally have lower metallicities. Specifically, for normal star-forming galaxies with stellar masses between $10^{8.5}$ and $10^{10.5}$ M$_{\odot}$, \citet{Tremonti04} find the correlation between stellar mass and metallicity to be roughly linear. Owing to the small coverage in stellar mass space, our LSBG sample is not sufficient for fitting a certain $M$--$Z$ relation, but galaxies in our sample are consistent with the relatively low abundances for their stellar masses.

\subsection{The Gas Mass Fraction vs. Oxygen Abundance}
The gas consisting mostly of hydrogen and helium is the raw material out of which stars and metals are formed. In LSBGs, the largest constituent of gas is assumed to be neutral atomic hydrogen. The molecular gas remains unknown in quantity but, fortunately, it is not expected to contribute greatly to the total mass of gas in LSBGs due to their low metallicities and low dust content \citep{Schombert90,de Blok98_2}. As for ionized gas, it contributes negligibly to the mass of LSBGs \citep{Schombert01}.

The gas mass fraction ($f_{g}$) refers to the fraction of baryons in gaseous form. 
It can be simply measured by the ratio of the total mass in gas ($M_{\rm gas}$, including molecular, atomic, ionized, and metals) to the combined total mass in gas and stars ($M_{\star}$):

\begin{equation}
f_{g}=\frac{M_{\rm gas}}{M_{\rm gas}+M_{\star}} .
\end{equation}

 So, taking the contribution of helium and metals into account, the total gas mass of LSBGs in units of solar mass can be given by  
 
\begin{equation}
M_{\rm gas}=\frac{M_{\rm H~I}}{X},
\end{equation}
\noindent
where $X$ is the fraction of the gas mass in the form of hydrogen, adopted to be 0.733 \citep{de Blok98_1}.

Following the formula above, $M_{\rm gas}$ for our edge-on LSBG sample can be easily computed by multiplying $M_{\rm H~{\sc{i}}}$ (Table ~\ref{tbl-1}) which is available from the $\alpha.$40 catalog \citep{Haynes11} by 1/0.733.
For $M_{\star}$,we have previously derived two sets of stellar masses for our LSBG sample in \S 5.3.1 (Table ~\ref{tbl-4}). Consequently, we have two sets of gas mass fraction, $f_{g}$, for our sample. 

\begin{figure*}
 \centering
 \includegraphics[scale=0.60]{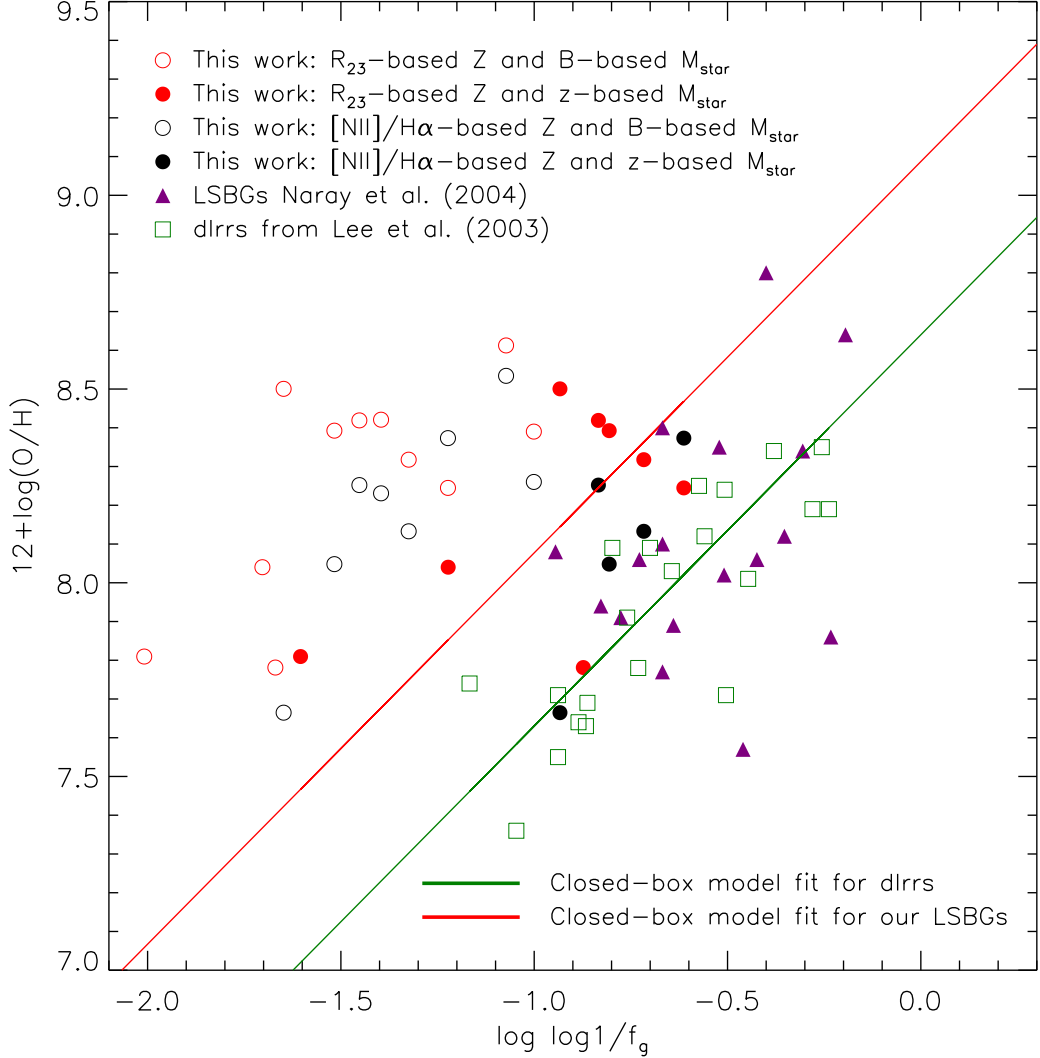}%fig_metal2.pro(linux)
 \caption[]{Gas mass fraction vs. metallicity for LSBGs. Different sets of stellar masses and metallicities are shown for our LSBG sample. The open and filled circles show stellar masses respectively estimated from the $B$-band method and the SDSS SED-fitting method, while the black and red circles respectively represent the $R_{23}$-based and [N~{\sc{ii}}]/H$\alpha$-based metallicities. Purple triangles and dark-green squares are respectively the LSBG sample of \citet{de Naray04} and the dIrr sample of \citet{Lee03}. The dark-green solid line with a slope of unity indicates the behavior of the closed-box model of chemical evolution with an oxygen yield of $2.22 \times 10^{-3}$ for the dIrr sample of \cite{Lee03}. The red solid line is the `hypothetical' closed-box model fit for our LSBG sample with a different oxygen yield of $6.23\times10^{-3}$. \label{fig:gas_metal}}
\end{figure*}

In Figure ~\ref{fig:gas_metal}, we show oxygen abundance versus inverse gas mass fraction for our LSBG sample. Both sets of gas mass fractions are shown for our sample. The filled circles represent for$f_{g}$ being computed from the ``$z$-based ''$M_{\star}$ while the open circles are for $f_{g}$ being computed from the ``$B$-based'' $M_{\star}$. Besides, two sets of metallicities are also presented  with red circles showing $R_{23}$-based metallicities and black circles showing [N~{\sc{ii}}]/H$\alpha$-based metallicities. For comparisons, the LSBG sample from \citet{de Naray04} (purple triangles) and the dIrr sample from \citet{Lee03} (green squares) are also plotted. 

For our LSBG sample, data points with ``$B$-based'' stellar mass (red and black open circles) or with ``$z$-based'' stellar mass but $R_{23}$-based metallicities (red filled circles) have systematically biased to much larger gas mass fraction, compared with other samples (e.g., the LSBG sample from \citealt{de Naray04}; the dIrr sample from \citealt{Lee03}). However, data points with ``$z$-based'' stellar mass and [N~{\sc{ii}}]/H$\alpha$-based metallicities (black filled circles) show general consistency with other samples in the common high-$f_{g}$ region. The systematically larger gas mass fractions are more likely to be caused by different methods used for estimating stellar mass and metallicity. Again, we prefer the ``$z$-based'' stellar masses for our sample because LSBGs are acknowledged to be dominated by low-mass stars that would radiate light predominantly at red and near-infrared wavelengths. Using $B$-band luminosity method is very likely to underestimate stellar masses of LSBGs and then overestimate their gas mass fractions. More details on this issue will be discussed in \S 6.2. 
 
\section{Discussion}
\subsection{Stellar Populations}
\subsubsection{Coarse Stellar Populations Results }
Stellar populations are the fossils of the past star formation and evolution of galaxies. However, the sky background makes it very difficult to study the faint stellar continuum and absorption lines in the entire LSBG unless very large telescopes and extraordinarily long exposure times are used. Because of such difficulties, most spectroscopic studies of LSBG stellar populations have been based on spectra of the brighter bulges (e.g., Morelli et al. 2012) or H~{\sc{ii}} regions. A slit of appropriate width and length positioned along the major axis can generally capture light from almost all regions of an edge-on galaxy, especially of a super-thin galaxy. Thus, the spectral properties may in principle be much better determined for edge-on LSBGs compared with face-on ones.

If we expect to derive reliable stellar populations with full-spectrum fitting stellar population synthesis method, we must use high-quality spectra, especially with well-calibrated spectral continuum shapes. Given the uncertainties associated with continuum shapes of the Palomar Hale spectra, we decided to use only the three Lick Shane spectra (of AGC~191770, AGC~228010, and AGC~253926) for our stellar-population study of LSBGs. All these three spectra have S/N $> 10$ in the continuum wavelength interval 4730--4780~\AA\ and meet our requirements.

For stellar population synthesis, STARLIGHT \citep{Cid Fernandes05} is a widely used tool. It can derive the age and metallicity of a galaxy, as well as its extinction and velocity dispersion, by fitting its observed spectrum with a superposition of populations of various ages and metallicities from stellar population models such as STARLIGHT \citep[][hereafter BC03]{Bruzual03}. We used STARLIGHT to perform stellar population synthesis for the integrated spectra of the three edge-on LSBGs observed at Lick. For the reddening, both the \citet[][hereafter CAL]{Calzetti00} law and the \citet[][hereafter CCM]{Cardelli89} law were separately used during the fit, although the problem of dust attenuation is minimal in LSBGs because they are acknowledged to be dust poor.

\begin{figure*}
 \centering
 \includegraphics[scale=0.23]{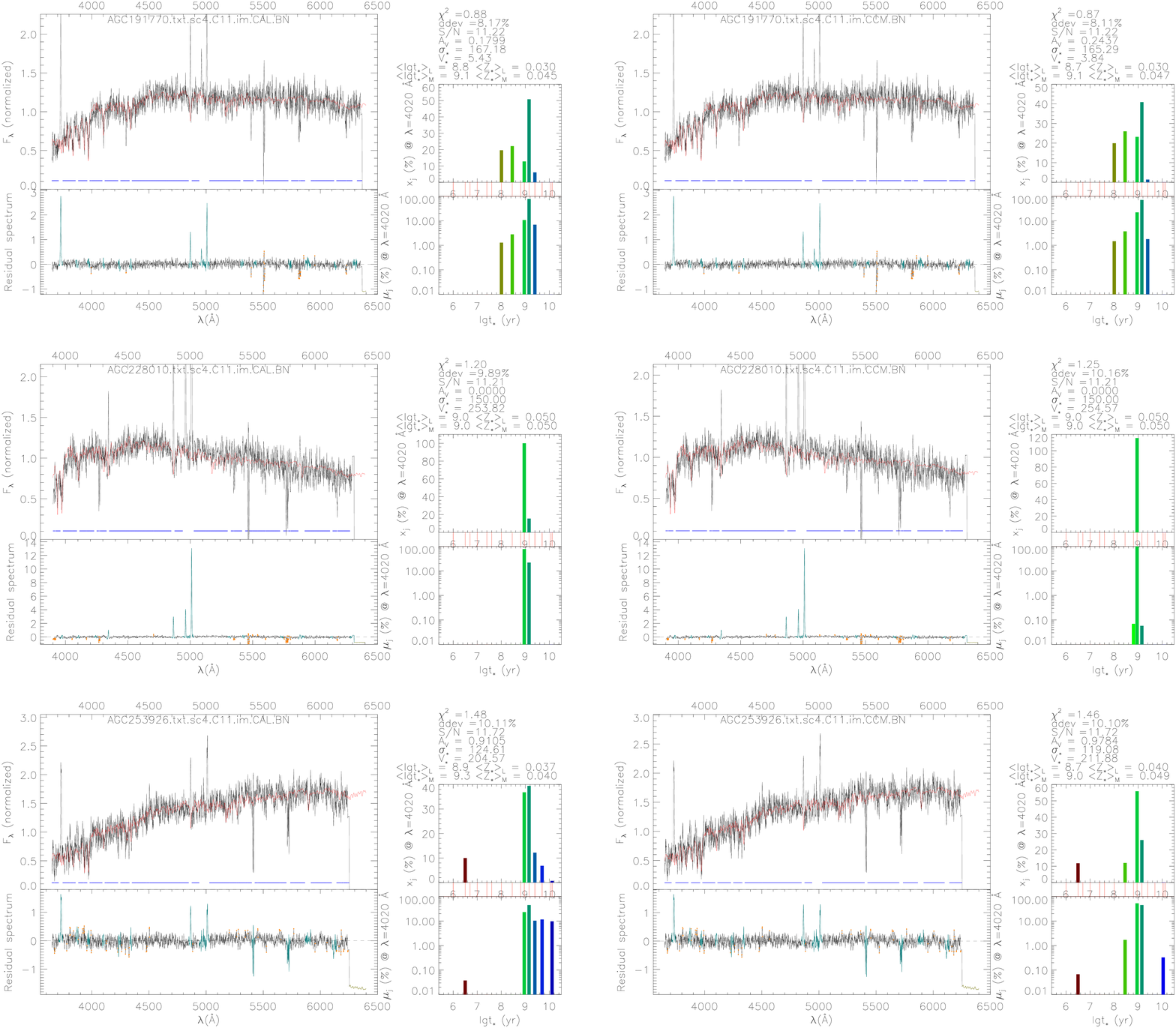}%call_starlightplot.pro
 \caption[]{STARLIGHT stellar population synthesis results for three edge-on LSBGs. The left and right columns are respectively for the CAL and CCM reddening laws. In each panel, the top-left part shows the observed spectrum (black) and the STARLIGHT fit (red). The blue line is the error spectrum, with gaps marking bad pixels. However, we have not extracted the error spectrum from our data, so the blue line here is simply a horizontal line of value 0. The bottom-left part shows the residual spectrum, with cyan corresponding to masked windows (e.g., emission lines) and yellow marking clipped pixels (e.g., bad sky subtraction). The plots at right show age distributions of the resulting stellar populations expressed as a light fraction (top) and a mass fraction (bottom). The bar-code panel (between the top-right and bottom-right parts) indicates the ages of all stellar populations included during the fit. \label{fig:starlight}}
\end{figure*}

We show the fit plot in Figure ~\ref{fig:starlight}. As said by \citet{Cid Fernandes05}, the individual stellar population components from STARLIGHT are very uncertain because the existence of multiple solutions is an old known problem in stellar population synthesis. Thus, we should avoid focusing in detail on any individual stellar populations. Instead, we jump straight to results based on more robust descriptions of the resultant stellar populations. As suggested in the STARLIGHT paper \citep{Cid Fernandes05}, a coarse but robust description of the star-formation history of our galaxies may be obtained by binning the resultant individual stellar populations onto ``young"  (Y; age $< 1$~Gyr) and ``old '' (O; age $> 1$~Gyr) components. In order to show the respective contributions of the two components, we provide the ``light fraction'' (LF) out of the total light and the ``mass fraction'' (MF) out of the total mass for Y and O components in Table ~\ref{tbl-5}. Results from both reddening laws (CAL and CCM) for each LSBG seem to be consistent. Qualitatively, for both light and mass, the contributions from the O components seem to be much larger than those  from the Y components for these three LSBGs. Thus, rather than a quantitative conclusion, we can draw a qualitative conclusion that these three LSBGs may have a potential dearth of stars with ages $<1$~Gyr. Instead, they are more likely to be dominated by stars with ages $>1$~Gyr.  This supports the general notion that LSBGs have a very low level of current star-formation activity. 

Comparing with the previously measured subsolar gas-phase metallicities for the first three galaxies in Table~\ref{tbl-3}, their stellar metallicities (from stellar population synthesis; Table~\ref{tbl-5}) are about solar. On one hand, this inconsistency is likely to be caused mainly by the large uncertainties of the stellar metallicities. As can be deduced from the amount of work devoted to this topic of stellar population synthesis, recovering the stellar content of a galaxy with accurately quantitative ages and metallicities from its observed integrated spectrum is not an easy task. It needs a high-quality integrated spectrum, as well as models which are made and calibrated to mimic nature to the greatest degree. Any stellar population model, such as STARLIGHT (BC03), is based on a stellar library. Usually, the stellar library does not sample a sufficiently wide range in metallicity, making stellar metallicity notoriously more difficult to access than other properties. Also, to get the metallicity correct, one must have correct ages, as they affect one another in these integrated spectral fits. Since our derived stellar ages have large uncertainties, the stellar metallicities are necessarily also poorly constrained.  So, stellar metallicities determined here from stellar population synthesis (Table~\ref{tbl-5}) carry large uncertainties, and are not reliable as specific values for comparisons and scientific studies; the gas-phase metallicities (Table~\ref{tbl-3}) from the emission-line flux ratios are much more reliable.

\subsubsection{Effect of Blue Light Loss on Conclusions}
We use the blue-side spectra for studying stellar populations by full-spectrum fitting, so the accuracy of the relative flux calibration becomes an important factor for the reliability of the stellar population results. If the slit is not positioned along the atmospheric dispersion, the light loss at blue-end wavelengths would be greater than that at red-end wavelengths. As shown in Table~\ref{tbl-2} in Section 3.2, if a star was being centered at 6000~\AA\ and observed with the same slit positioning schemes and observing configurations as those in our program, it will be off center by 0.61$\arcsec$ at 3600~\AA\ and 0.05~$\arcsec$ in the opposite direction at 6500~\AA\ for AGC~191770, by 0.8$\arcsec$ at 3600~\AA\ and 0.06$\arcsec$ in the opposite direction at 6500~\AA\ for AGC~228010, and  by 0.8$\arcsec$ at 3600~\AA\ and 0.06$\arcsec$ in the opposite direction at 6500~\AA\ for AGC~253926.  The light loss at 3600~\AA\ is (respectively) 4.8\%, 15.2\%, and 0.6\% greater than that at 6500~\AA\ for AGC~191770, AGC~228010, and AGC~253926.

Obviously, the relative blue-light loss of AGC~228010 is the largest among the three galaxies. However, such an amount of light loss is calculated under the assumption that our targets were stars. Actually, as seen in Figure ~\ref{fig:edgeon-slit}, they are galaxies with extended areas larger than the slit width. So, the lost blue light is partly gained back from the neighboring pixels, and the effect on relative flux calibration of blue-end spectra is not that significant, although it still exists. The probable light loss at blue-end wavelengths ($\sim3600$~\AA) would cause an underestimate of late A-type stars. Even with some blue-light loss, we could still obtain the same qualitative conclusion that these three LSBGs have a potential dearth of stars with ages $<1$~Gyr because late A-type stars typically have ages older than $\sim1$~Gyr.

\subsubsection{Limitations Caused by Very Small Sample}
Limited by the reliability of a spectral continuum, the number of spectra that we can use for a possibly reliable conclusion from full-spectrum fitting is only three. Undoubtedly, this is from a representative sample of the whole edge-on LSBG sample, and not even representative of our final 12 observing targets. Thus, conclusions from the three galaxies may not apply more generally to LSBGs. To get more universal conclusions on stellar populations of LSBGs, we will seek sufficient observing time in the future, obtaining stellar continuum spectra of good quality for a larger and more representative sample of edge-on LSBGs.

\begin{deluxetable}{ccccccc}
\tabletypesize{\scriptsize}
\tablecaption{Stellar Population Synthesis Results for Three Edge-on LSBGs \label{tbl-5}}
\tablewidth{0pt}
\tablehead{
\colhead{Parameter} & \colhead{} & \colhead{} & \colhead{} & \colhead{Galaxy Name} &\colhead{} & \colhead{}  \\

\colhead{} & \colhead{~~~~~~AGC 191770} & \colhead{} & \colhead{~~~~~~AGC 228010} & \colhead{} &\colhead{~~~~~~AGC 253926} & \colhead{} \\

\colhead{} & \colhead{CAL} & \colhead{CCM} & \colhead{CAL} & \colhead{CCM} & \colhead{CAL} & \colhead{CCM} 

}
\startdata
log(age$_{L}$) & 8.8 &8.7 &9.0 &9.0 & 8.9 & 8.7\\
log(age$_{M}$) & 9.1 &9.1 &9.0 &9.0 & 9.3 & 9.0\\
$Z_{L}$\tablenotemark{a}&0.030 &0.030 &0.050 &0.050 &0.037 &0.040\\
$Z_{M}$&0.045 &0.047 &0.050 &0.050 &0.040 &0.049\\
$A_{V}$ (mag) & 0.18 & 0.24 & 0.00 & 0.00 & 0.91 & 0.99\\
\hline
LF$_{\rm YSP}\tablenotemark{b}$ & 0.0 &0.0 &0.0 &0.0 &9.4 &11.2\\
%LF$_{\rm ISP}$ & 94.6 &98.8 &100 &100 &72.0 &88.8\\
%LF$_{\rm OSP}$ & 5.4 & 1.2 &0.0 &0.0 &18.6 & 0\\
LF$_{\rm OSP}$ & 100.0 & 100.0 &100.0 &100.0 &90.6 & 88.8\\
\hline
MF$_{\rm YSP}$\tablenotemark{c} & 0.0 & 0.0 &0.0 & 0.0 &0.04 &0.1\\
%MF$_{\rm ISP}$ & 93.05 &98.21 &100 &100.0 &68.20 &99.6\\
%MF$_{\rm OSP}$ & 6.95 & 1.79 & 0.0 &0.0 &31.76 &0.3\\ 
MF$_{\rm OSP}$ & 100.0 & 100.0& 100.0 &100.0 &99.96 &99.9\\
\hline
\enddata
%% Text for table notes should follow after the \enddata but before
%% the \end{deluxetable}. Make sure there is at least one \tablenotemark
%% in the table for each \tablenotetext.
%\tablecomments{Table \ref{tbl-2} is published in its entirety in the 
%electronic edition of the {\it Astrophysical Journal}.  A portion is 
%shown here for guidance regarding its form and content.}
\tablenotetext{a}{$Z$ refers to the mass fraction of all ``metals" of a star. $Z_{L}$ and $Z_{M}$ refer to metallicities of luminosity-weighted and mass-weighted stellar populations, respectively. }
\tablenotetext{b}{ LF refers to light fraction to the total light. }
\tablenotetext{c}{MF refers to mass fraction to the total stellar mass. }
\end{deluxetable}

\subsection{Stellar Mass}

In Figure~\ref{fig:gas_metal}, the gas mass fractions computed from two different sets of stellar masses for our sample have obvious discrepancies. The gas mass fractions from the ``$B$-based'' stellar masses are apparently higher than those from the ``$z$-based" stellar masses. 

Which set of stellar mass is more appropriate for LSBGs? Here, we prefer the ``$z$-based'' stellar masses because of the following reasons: (1) The luminosity at relatively longer wavelengths (the red or near-infrared bands) is more ideal for tracing the stellar mass because it is less affected by extinction and other factors that affect shorter-wavelength bands. (2) LSBGs have few blue, massive stars formed in the past 1~Gyr; the stellar mass is predominantly in low mass stars which radiate most of their light at red and near-infrared wavelengths. We checked the {\it WISE} satellite 3.4~$\mu$m and 4.6~$\mu$m images of our LSBGs and found that all of them are detected in both bands, especially in the bulge regions. Thus, the stellar masses of LSBGs would be underestimated from their $B$-band luminosity. (3) The ``$z$-based" stellar masses of our LSBG sample are $M_{\star} \approx 10^{8.0}$--$10^{10.0}$ M$_{\odot}$, which agrees with the conclusion of other studies that LSBGs are mostly comparable in mass with prominent spirals that define the Hubble sequence \citep[e.g.,][]{McGaugh94}.

%\begin{figure*}
% \centering
 %\includegraphics[scale=0.40]{fig7.eps}%fig_images.pro(mac)
 %\caption[]{Images of our 12 edge-on LSBGs in the {\it WISE} 3.4~$\mu$m band. The size of each panel is $100\arcsec \times 100\arcsec$. In each panel, North is up and East is to the left.\label{fig:wise_images}}
%\end{figure*}
  
\subsection{Chemical Evolution of LSBGs}
In the closed-box model of the chemical evolution of galaxies, the initially metal-free gas is permanently locked into stars with neither inflow nor outflow and only evolves ``in isolation'' with time. Other assumptions include an invariant stellar initial mass function and instantaneous recycling. The relation between the metallicity ($Z$) and the gas mass fraction ($f_{g}$) for a system that evolves in isolation is

\begin{equation}
Z=y\,{\rm ln}(1/f_{g}),
\end{equation}
\noindent
where $y$ is the ``yield,'' the ratio of the mass of newly formed metals to the mass of gas permanently locked in stars. Here, $Z$ is best expressed as an elemental abundance in logarithmic form. For oxygen, following \citet{Lee03}, this equation can be rewritten as 

\begin{eqnarray}
{\rm log}\,Z_{\rm O} = {\rm log}\,y_{\rm O} + {\rm log}\,{\rm ln}(1/f_{g}),\\
12 + {\rm log}\,{\rm (O/H)} = 12 + {\rm log}\,(0.196y_{\rm O}) + {\rm log}\,{\rm log}(1/f_{g}).
\end{eqnarray}
\noindent
In a plot of oxygen abundance, 12 + log(O/H), against inverse gas mass fraction, conveyed by log\,log($1/f_{g}$), the closed-box model predicts a slope of unity. Deviations from this slope may be an indication for the inflow or outflow of gas. The oxygen yield, $y_{\rm O}$, can be derived from the intercept of the plot.

Following \citet{Lee03}, for the field dIrrs, there is an excellent correlation represented by a line with a slope of unity (green solid line in Figure ~\ref{fig:gas_metal}) which is consistent with the closed-box model of the chemical evolution of galaxies. The standard deviation of the fit is 0.126 dex in log(O/H). From the intercept of the line, the effective oxygen yield by mass ($y_{\rm O}$) for the field dIrrs from \citet{Lee03} is 2.22$\times$10$^{-3}$. Obviously, the LSBG data (purple open triangles) from \citet{de Naray04} are generally consistent with the field dIrr data (green open squares) and can also be fitted by the closed-box model fit (green solid line) of \citet{Lee03}, albeit with more scatter. 

We cannot tell for sure whether our LSBG sample can be explained with the closed-box model by fitting; our sample is too small and also too narrow in the coverage of both gas mass fraction and gas-phase metallicity to show a true relation, even if some essential relations do indeed exist. However, to compare with the field dIrrs from \citet{Lee03} and the LSBGs from \citet{de Naray04}, we still forced our LSBG data (all filled circles with ``$z$-based'' stellar mass) to be fitted by a line with a slope of unity (red solid line). The standard deviation of this fit is 0.27 dex in log(O/H).  From this artificial fitting, the effective oxygen yield by mass for our LSBG sample is $6.23 \times 10^{-3}$ under the assumption of a closed-box model. The value of true yield is often considered to be the solar oxygen abundance by mass. For the Sun, the oxygen abundance is $12 +$ log(O/H) $=8.87$~dex \citep{Grevesse96,Lee03}, which is $9.00 \times 10^{-3}$ by mass.  Apparently, the effective oxygen yield for our LSBG sample is larger than that for the dIrr sample, and is 69\%  of the value for the true oxygen yield. The difference between the effective and true yields roughly represents the fraction of metal lost (up to 30\%) up to now \citep{Garnett02}. 

Such a value of metal loss can be reasonable since it is often argued that LSBGs have shallow potential wells, making it easier for mass loss to occur. However, given the substantial uncertainties in both metallicity and gas mass fraction for our LSBG sample, this mass-loss value of 30\% is only a rough estimate, and we cannot completely rule out a closed-box scenario. Additionally, judging from the offset between the two closed-box lines in Figure~\ref{fig:gas_metal}, we find evidence that our LSBGs retain up to $\sim 3$ times as much of their metals compared with dwarf irregulars (green squares), consistent with metal retention being related to galaxy mass. In any case, although the closed-box scenario cannot be completely ruled out for our LSBG sample, there is still substantial room for considering other, more complicated chemical evolution models with outflow or even infall \citep[e.g.,][]{Edmunds90,Chang10} for our LSBG sample.

\subsection{Are LSBGs a Fundamentally Different Type?}
In the $z$-based stellar masses (Table ~\ref{tbl-4}), our LSBG sample is estimated to be $10^{8.3}$--$10^{10.0}$ M$_{\odot}$, showing no fundamental differences from prominent spiral galaxies. Furthermore, comparing with previously published luminosity-metallicity plots for LSBGs and other galaxy types (Figure~\ref{fig:lumi_metal}), our LSBG sample agrees well with the LSBG sample of \citet{de Naray04} and also generally the sample of normal  galaxies. In the mass-metallicity plot (Figure~\ref{fig:mass_metal}), our LSBG sample is also consistent with the LSBG sample of \citet{de Naray04} and with the relatively low abundance for their stellar mass, which is appropriate for normal galaxies. Therefore, from the behavior in the $L$--$Z$ and $M$--$Z$ plots, our LSBG sample does not seem to be fundamentally different from normal disk galaxies.  

Regarding the plot of gas mass fraction versus oxygen abundance (Figure~\ref{fig:gas_metal}), we prefer the $z$-based stellar masses (black or red filled circles). Our LSBG sample is not likely to be explained by simple closed-box models, although the galaxies evolve in relative isolation. Systematically biased to larger gas mass fractions, our LSBG sample has higher oxygen abundances for smaller gas mass fractions, which is also shown by other samples. Such qualitative agreement in gas mass fraction versus metallicity distribution makes the LSBGs seem consistent with other galaxy types.

We therefore suggest that LSBGs are probably not fundamentally different from normal galaxies, but are just the continuous extension to the lower surface-brightness end of normal disk galaxies. 

\section{Conclusions}

We have obtained long-slit spectra of 12 edge-on LSBGs selected from the EGIS-$\alpha.$40 sample.  Observations were made with the Kast spectrograph on the 3-m Shane telescope at Lick Observatory for 3 galaxies and with the Double Spectrograph on the 5-m Hale telescope at Palomar Observatory for 9 galaxies. We adopted a median-filter method to estimate the sky background of objects observed at Lick Observatory and a second-order polynomial to fit the sky background of objects observed at Palomar Observatory, which can be comparable to or even exceed the galaxy surface brightnesses.

The stellar populations of three of the LSBGs were studied with the program STARLIGHT, as their spectral quality is sufficiently high. We find that 
our LSBGs are dominated by stellar populations with ages $>1$~Gyr and have a potential dearth of stars with ages $<1$~Gyr. This indicates that LSBGs probably have a very low level of star-formation activity over the past 1~Gyr, which may be the most likely explanation for their low surface brightness.

We have measured oxygen abundances (12 + log(O/H)) using both the $R_{23}$ and [N~{\sc{ii}}]/H$\alpha$ diagnostics after performing reddening corrections for the observed fluxes of emission lines. The $R_{23}$-based metallicities are 8.00--8.62, with a mean of 8.30 dex and a median of 8.26 dex. The [N~{\sc{ii}}]/H$\alpha$-based metallicities are 7.67--8.54, with a mean of 8.20 dex and a median of 8.26 dex. The two sets of oxygen abundances are consistent with each other. Comparing with the luminosity-metallicity relation established in previous work on LSBGs, normal spirals, and other galaxy types, our LSBGs tend to be consistent with normal galaxies.  

We derive two sets of stellar masses for our LSBG sample. One set is  the $z$-based stellar masses which are $10^{8.0}$--$10^{10.0}$ M$_{\odot}$, and the other set is the $B$-based stellar masses which are $10^{7.5}$--$10^{9.5}$ M$_{\odot}$, 0.5 order of magnitude lower. We argue that the $z$-based stellar masses are better for our LSBG sample because such galaxies primarily consist of low mass stars which radiate their light mostly at red and infrared wavelengths. Stellar masses derived from the $B$ band might be underestimated. According to the $z$-based stellar masses, our LSBGs are comparable to normal spiral galaxies in stellar mass, but they have considerably lower abundances.     

In the plot of gas mass fraction versus oxygen abundance, we find that like other galaxy types, our LSBGs with relatively high ``$z$-based'' stellar masses have lower metallicities for higher gas mass fractions. Regarding the chemical evolution of our sample, the LSBG data appear to allow for up to 30\% metal loss, but we cannot completely rule out the closed-box model. Additionally, we find evidence that our galaxies retain up to about 3 times as much of their metals compared with dwarfs, consistent with metal retention being related to galaxy mass.

Combining the above conclusions, our LSBG sample is generally consistent with the luminosity-metallicity, mass-metallicity, and gas mass fraction-metallicity plots of normal spiral galaxies and other galaxy types (e.g., dIrrs, starbursts). Thus, we speculate that LSBGs are probably not fundamentally different from normal galaxies, but rather are just the continuous extension in surface brightness to the low end.

%% If you wish to include an acknowledgments section in your paper,
%% separate it off from the body of the text using the \acknowledgments
%% command.

%% Included in this acknowledgments section are examples of the
%% AASTeX hypertext markup commands. Use \url without the optional [HREF]
%% argument when you want to print the url directly in the text. Otherwise,
%% use either \url or \anchor, with the HREF as the first argument and the
%% text to be printed in the second.

\acknowledgments

We acknowledge the anonymous referee for whose detailed comments helped strengthen this paper. This research uses data obtained through the Telescope Access Program (TAP), which has been funded by the National Astronomical Observatories of China, the Chinese Academy of Sciences (the Strategic Priority Research Program "The Emergence of Cosmological Structures" Grant No. XDB09000000), and the Special Fund for Astronomy from the Ministry of Finance. Three observational nights (20--22 Sep. 2015) with the Double Spectrograph on the 5-m Hale telescope were distributed to us for scientific studies of LSBGs via the TAP. Observations obtained with the Hale Telescope at Palomar Observatory were obtained as part of an agreement between the National Astronomical Observatories, Chinese Academy of Sciences, and the California Institute of Technology. Additionally, we are grateful to the staffs of Lick and Palomar Observatories for their assistance with the observations. Research at Lick Observatory is partially supported by a generous gift from Google. W. Du thanks Cheng Cheng, Zhimin Zhou and Fan Yang in NAOC for discussions concerning data reduction, and thanks Ruixiang Chang in SHAO for helpful suggestions during revision.

This work is supported by the National Natural Science Foundation of China (Grant Nos. 11403037, 11225316, and 11173030),  the Strategic Priority Research Program ``The Emergence of Cosmological Structures" of the Chinese Academy of Sciences (Grant No. XDB09000000), and the Key Laboratory of Optical Astronomy, NAOC. A.V.F. has received generous financial assistance from the Christopher R. Redlich Fund, the TABASGO Foundation, and NSF grant AST-1211916. 

%% To help institutions obtain information on the effectiveness of their
%% telescopes, the AAS Journals has created a group of keywords for telescope
%% facilities. A common set of keywords will make these types of searches
%% significantly easier and more accurate. In addition, they will also be
%% useful in linking papers together which utilize the same telescopes
%% within the framework of the National Virtual Observatory.
%% See the AASTeX Web site at http://aastex.aas.org/
%% for information on obtaining the facility keywords.

%% After the acknowledgments section, use the following syntax and the
%% \facility{} macro to list the keywords of facilities used in the research
%% for the paper.  Each keyword will be checked against the master list during
%% copy editing.  Individual instruments or configurations can be provided 
%% in parentheses, after the keyword, but they will not be verified.

{\it Facilities:} \facility{Palomar Hale (DBSP)}, \facility{Lick Shane (Kast)}

\clearpage

%% Use the figure environment and \plotone or \plottwo to include
%% figures and captions in your electronic submission.
%% To embed the sample graphics in
%% the file, uncomment the \plotone, \plottwo, and
%% \includegraphics commands
%%
%% If you need a layout that cannot be achieved with \plotone or
%% \plottwo, you can invoke the graphicx package directly with the
%% \includegraphics command or use \plotfiddle. For more information,
%% please see the tutorial on "Using Electronic Art with AASTeX" in the
%% documentation section at the AASTeX Web site, http://aastex.aas.org/
%%
%% The examples below also include sample markup for submission of
%% supplemental electronic materials. As always, be sure to check
%% the instructions to authors for the journal you are submitting to
%% for specific submissions guidelines as they vary from
%% journal to journal.

%% This example uses \plotone to include an EPS file scaled to
%% 80% of its natural size with \epsscale. Its caption
%% has been written to indicate that additional figure parts will be
%% available in the electronic journal.

\clearpage

%% Here we use \plottwo to present two versions of the same figure,
%% one in black and white for print the other in RGB color
%% for online presentation. Note that the caption indicates
%% that a color version of the figure will be available online.
%%

%% This figure uses \includegraphics to scale and rotate the still frame
%% for an mpeg animation.

%% If you are not including electonic art with your submission, you may
%% mark up your captions using the \figcaption command. See the
%% User Guide for details.
%%
%% No more than seven \figcaption commands are allowed per page,
%% so if you have more than seven captions, insert a \clearpage
%% after every seventh one.

%% Tables should be submitted one per page, so put a \clearpage before
%% each one.

%% Two options are available to the author for producing tables:  the
%% deluxetable environment provided by the AASTeX package or the LaTeX
%% table environment.  Use of deluxetable is preferred.
%%

%% Three table samples follow, two marked up in the deluxetable environment,
%% one marked up as a LaTeX table.

%% In this first example, note that the \tabletypesize{}
%% command has been used to reduce the font size of the table.
%% We also use the \rotate command to rotate the table to
%% landscape orientation since it is very wide even at the
%% reduced font size.
%%
%% Note also that the \label command needs to be placed
%% inside the \tablecaption.

%% This table also includes a table comment indicating that the full
%% version will be available in machine-readable format in the electronic
%% edition.

\clearpage

%% If you use the table environment, please indicate horizontal rules using
%% \tableline, not \hline.
%% Do not put multiple tabular environments within a single table.
%% The optional \label should appear inside the \caption command.

\clearpage

%% If the table is more than one page long, the width of the table can vary
%% from page to page when the default \tablewidth is used, as below.  The
%% individual table widths for each page will be written to the log file; a
%% maximum tablewidth for the table can be computed from these values.
%% The \tablewidth argument can then be reset and the file reprocessed, so
%% that the table is of uniform width throughout. Try getting the widths
%% from the log file and changing the \tablewidth parameter to see how
%% adjusting this value affects table formatting.

%% The \dataset{} macro has also been applied to a few of the objects to
%% show how many observations can be tagged in a table.

\clearpage

%% Tables may also be prepared as separate files. See the accompanying
%% sample file table.tex for an example of an external table file.
%% To include an external file in your main document, use the \input
%% command. Uncomment the line below to include table.tex in this
%% sample file. (Note that you will need to comment out the \documentclass,
%% \begin{document}, and \end{document} commands from table.tex if you want
%% to include it in this document.)

%% \input{table}

%% The following command ends your manuscript. LaTeX will ignore any text
%% that appears after it.


\begin{thebibliography}{}
\bibitem[Abazajian et al.(2009)]{Abazajian09} Abazajian K. N., Adelman-McCarthy J. K., Agueros M. A. et al. 2009, \apjs, 182, 543
%\bibitem[Allende Prieto et al. (2001)]{Allende Prieto01} Allende Prieto C., Lambert L. D., Asplund M. 2001, \apj, 556, L63 
\bibitem[Bell et al.(2003)]{Bell03} Bell E. F., McIntosh D. H., Katz N., Weinberg M. D. 2003, \apjs, 149, 289
\bibitem[Bizyaev et al.(2014)]{Bizyaev14} Bizyaev D. V., Kautsch S. J., Mosenkov A. V. et al. 2014, \apj, 787, 24
\bibitem[Bothun et al.(1987)]{Bothun87} Bothun G. D., Impey C. D., Malin D. F., Mould J. R. 1987, \aj, 94, 23
\bibitem[Bothun et al.(1997)]{Bothun97} Bothun G. D. et al. 1997, \pasp, 109, 745
\bibitem[Brodie $\&$ Huchra(1991)]{Brodie91} Brodie J. P., Huchra J. P. 1991, \apj, 379, 157
\bibitem[Bruzual $\&$ Charlot(2003)]{Bruzual03} Bruzual G., Charlot S. 2003, \mnras, 344, 1000
\bibitem[Burkholder et al.(2001)]{Burkholder01} Burkholder V., Impey C., Sprayberry D. 2001, \aj, 122, 2318
\bibitem[Calzetti et al.(2000)]{Calzetti00} Calzetti D., Armus L., Bohlin R. C. et al. 2000, \apj, 533, 682
\bibitem[Cardelli et al.(1989)]{Cardelli89} Cardelli J. A., Clayton G. C., Mathis J. S. 1989, \apj, 345, 245
\bibitem[Ceccarelli et al.(2012)]{Ceccarelli12} Ceccarelli L., Herrera-Camus R., Lambas D. G., Galaz G., Padilla N. D. 2012, \mnras, 426, 6
\bibitem[Chang et al.(2010)]{Chang10} Chang R. X., Hou J. L., Shen S. Y., Shu C. G. 2010, \apj, 722, 380
\bibitem[Cid Fernandes et al.(2005)]{Cid Fernandes05} Cid Fernandes R., Mateus A., Sodre L., Stasinska G., Gomes J. M. 2005, MNRAS, 358, 363
\bibitem[Cole et al.(2001)]{Cole01} Cole S., Norberg P., Baugh C. M. et al. 2001, \mnras, 326, 255
\bibitem[de Grijs(1998)]{de Grijs98} de Grijs R. 1998, MNRAS, 299, 595
\bibitem[de Blok(1998)]{de Blok98_1} de Blok W. J. G., McGaugh S. S. 1998, \apj, 508, 132
\bibitem[de Blok(1998)]{de Blok98_2} de Blok W. J. G., van der Hulst J. M. 1998, A$\&$A, 336, 49
\bibitem[de Naray(2004)]{de Naray04} de Naray R., McGaugh S., de Blok W. 2004, \mnras, 355, 887
\bibitem[Das et al.(2009)]{Das09} Das M. et al. 2009, \apj, 693, 1300
\bibitem[Denicolo et al.(2002)]{Denicolo02} Denicolo G., Terlevich R., Terlevich E., 2002, \mnras, 330, 69
\bibitem[Disney (1976)]{Disney76} Disney M. J. 1976, Nature, 263, 573
\bibitem[Du et al.(2015)]{Du15} Du W., Wu H., Lam M.-I. et al. 2015, \aj, 149, 199
\bibitem[Edmunds (1990)]{Edmunds90} Edmunds M. G. 1990, \mnras, 246, 678
\bibitem[Edmunds (1999)]{Edmunds99} Edmunds M. G. 1999, in Davies J. I., Impey C., Phillips S., ed., ASP Conf. Ser. Vol.170, The Low Surface Brightness Universe. Astron. Soc. Pac., San Francisco, p. 383
\bibitem[Faber(1973)]{Faber73} Faber S. M. 1973, \apj, 179, 731
\bibitem[Filippenko (1982)]{Filippenko82} Filippenko A. V. 1982, \pasp, 94, 715
\bibitem[Galaz et al.(2011)]{Galaz11} Galaz G. et al. 2011, \apj, 728, 74
\bibitem[Garnett et al.(1987)]{Garnett87} Garnett D. R., Shield G. A. 1987, \apj, 317, 82
\bibitem[Garnett (2002)]{Garnett02} Garnett R. D. 2002, \apj, 581, 1019
\bibitem[Grevesse et al. (1996)]{Grevesse96} Grevesse N., Noels A., Sauval A. J. 1996, in ASP Conf. Ser. 99, Cosmic Abundances: Proceedings of the 6th Annual October Astrophysics Conference, ed. S. S. Holt $\&$ G. Sonneborn (San Francisco: ASP), 117
\bibitem[Haberzettl et al.(2007)]{Haberzettl07} Haberzettl L. et al. 2007, A$\&$A, 471, 787
\bibitem[Haynes et al.(2011)]{Haynes11} Haynes M. P., Giovanelli R., Martin A. M. et al. 2011, \aj, 142, 170
\bibitem[Impey $\&$ Bothun(1997)]{Impey97} Impey C. $\&$ Bothun G. 1997, AR$\&$AA, 35, 267
\bibitem[Impey et al.(2001)]{Impey01} Impey C., Burkholder V., Sprayberry D. 2001, \aj, 122, 2341
\bibitem[Kauffmann et al.(2003)]{Kauffmann03} Kauffmann G., Heckman T. M., White S. D. M. et al. 2003, \mnras, 314, 33
\bibitem[Kelson (2003)]{Kelson03} Kelson D. D. 2003, \pasp, 155, 688
\bibitem[Kewley $\&$ Dopita (2002)]{Kewley02} Kewley L. J. \& Dopita M. A. 2002, \apjs, 142, 35
\bibitem[Kochanek et al.(2001)]{Kochanek01} Kochanek C. S., Pahre M. A., Falco E. E. et al. 2001, \apj, 560, 566
\bibitem[Lam et al.(2015)]{Lam15} Lam M. I., Wu H., Yang M. et al. 2015, \mnras, 446, 4291
\bibitem[Lamareille et al.(2004)]{Lamareille04} Lamareille F., Mouhcine M., Contini T., Lewis I., Maddox S. 2004, \mnras, 350, 396
\bibitem[Lee et al.(2003)]{Lee03} Lee H., McCall M. L.,Kingsburgh R. L., Ross R., Stevenson C. C. 2003, \aj, 125, 146
\bibitem[Lequeux et al.(1979)]{Lequeux79} Lequeux J., Peimbert M., Rayo J. F., Serrano A., Torres-Peimbert S. 1979, A$\&$A, 80, 155 
\bibitem[Liang et al.(2010)]{Liang10} Liang Y. C. et al. 2010, \mnras, 409, 213
\bibitem[Matthews $\&$ Gallagher(1997)]{Matthews97} Matthews L. D. $\&$ Gallagher J. S. \aj, 114, 1899
\bibitem[Matthews $\&$ Wood(2001)]{Matthews01} Matthews L. D. $\&$ Wood K. 2001, \apj, 548, 150
\bibitem[McGaugh(1991)] {McGaugh91} McGaugh S. S., 1991, \apj, 380, 140
\bibitem[McGaugh(1994)]{McGaugh94} McGaugh S. S. 1994, \apj, 426, 135
\bibitem[McGaugh(1996)]{McGaugh96} McGaugh S. S. 1996, \mnras, 280, 337
\bibitem[McGaugh $\&$ de Blok(1997)]{McGaugh97} McGaugh S. S., de Blok W. J. G. 1997, \apj, 481, 689
\bibitem[McGaugh $\&$ Bothun(1994)]{McGaugh94_1} McGaugh S. S. $\&$ Bothun G. D. 1994, \aj, 107, 530
\bibitem[Melbourne $\&$ Salzer(2002)]{Melbourne02} Melbourne J., Salzer J. J. 2002, \aj,123, 2302
\bibitem[Miller $\&$ Stone (1993)]{Miller93} {Miller} J. ~S. \& {Stone} R.~P.~S. 1993, {Lick Obs. Tech. Rep. 66} {Santa Cruz: Lick Obs.}
\bibitem[Minchin et al.(2004)]{Minchin04} Minchin R. F. et al. 2004, \mnras, 355, 1303
\bibitem[Morelli et al.(2012)]{Morelli12}Morelli L., Corsini E. M., Pizzella A. et al., 2012, \mnras, 423, 962
\bibitem[Oke \& Gunn(1982)]{okegunn82}Oke J.~B. \& Gunn J.~E. 1982, PASP, 94, 586 
\bibitem[O'Neil et al.(2000)]{O'Neil00} O'Neil K. et al. 2000, \aj, 119, 136
\bibitem[Osterbrock(1989)]{Osterbrock89} Osterbrock D. E. 1989, Astrophysics of Gaseous Nebulae and Active Galactic Nuclei (Mill Valley, CA: University Science Books) 
\bibitem[Pagel et al.(1979)]{Pagel79} Pagel B. E. J. et al. 1979, \mnras, 189, 95
\bibitem[Pettini $\&$ Pagel (2004)]{Pettini04} Pettini M. $\&$ Pagel B. E. J., 2004, \mnras, 348, L95
\bibitem[Schombert et al.(1990)]{Schombert90} Schombert J. M., Bothun G. D., Impey C. D., Mundy L. G. 1990, \aj, 100, 1523
\bibitem[Schombert et al.(2001)]{Schombert01} Schombert J., McGaugh S. S., Eder J. A. 2001, \aj, 121, 2420
\bibitem[Schombert et al.(1992)]{Schombert92} Schombert J. et al. 1992, \aj, 103, 1107
\bibitem[Schombert et al.(2013)]{Schombert13} Schombert J. et al. 2013, \aj, 146, 41
\bibitem[Skillman et al.(1989)]{Skillman89} Skillman E. D., Kennicutt R. C., Hodge P. W. 1989, \apj, 347, 875
\bibitem[Smith et al.(2002)]{Smith02} Smith J. A., Tucker D. L., Kent S., Richmond M. W. et al. 2002, \aj, 123, 2121
\bibitem[Trachternach et al.(2006)]{Trachternach06} Trachternach C., Bomans D. J., Haberzettl L., Dettmar R. J. 2006, A$\&$A, 458, 341
\bibitem[Tremonti et al.(2004)]{Tremonti04} Tremonti C. A., Heckman T. M., Kauffmann G. et al. 2004, \apj, 613, 898
\bibitem[van Dokkum et al.(2001)]{van Dokkum01} van Dokkum P. G. 2001, \pasp, 113, 1420
\bibitem[van der Hulst et al.(1993)]{van der Hulst93} van der Hulst J. M. et al. 1993, \aj, 106, 548
\bibitem[van der Kruit(1981)]{van der Kruit81} van der Kruit P. C. 1981, A$\&$A, 99, 298
\bibitem[van der Kruit(1988)]{van der Kruit88} van der Kruit P. C. 1988, A$\&$A, 192, 117
\bibitem[van der Kruit et al.(2001)]{van der Kruit01} van der Kruit P. C., Kregel M., Freeman C. K. 2001, A$\&$A, 379, 374
\bibitem[Wen et al.(2013)]{Wen13}Wen X. Q., Wu H., Zhu Y. N. et al. 2013, \mnras, 433, 2946
\bibitem[Wu et al.(2005)]{Wu05} Wu H., Cao C., Hao C. N. et al. 2005, \apj, 632, 79
\bibitem[Zaritsky et al.(1994)]{Zaritsky94} Zaritsky D., Kennicutt Jr. R. C., Huchra J. P. 1994, \apj, 420, 87
\bibitem[Zhu et al.(2010)]{Zhu10} Zhu Y. N., Wu H., Li H. N., Cao C. 2010, RAA, 10, 329
\end{thebibliography}
\end{document}